\documentclass[prd,nofootinbib,showpacs,preprint]{revtex4}
\usepackage{amsmath}
\usepackage{graphicx}
\graphicspath{{Figs/}}
\usepackage{dcolumn}
\usepackage{bm}
\usepackage{amssymb}
\usepackage[usenames,dvipsnames]{color}
\usepackage{slashed}
\usepackage[dvipdfm,colorlinks,citecolor=blue]{hyperref}
\begin{document}
\title{Type II Seesaw Origin of Non-zero $\theta_{13}, \delta_{CP} $ and Leptogenesis}
\author{Debasish Borah}
\email{dborah@tezu.ernet.in}
\affiliation{Department of Physics, Tezpur University, Tezpur-784028, India}


\begin{abstract}
We discuss the possible origin of non-zero reactor mixing angle $\theta_{13}$ and Dirac CP phase $\delta_{CP}$ in the leptonic sector from a combination of type I and type II seesaw mechanisms. Type I seesaw contribution to neutrino mass matrix is of tri-bimaximal (TBM) type which gives rise to vanishing $\theta_{13}$ leaving the Dirac CP phase undetermined. If the Dirac neutrino mass matrix is assumed to take the diagonal charged lepton type structure, such a TBM type neutrino mass matrix originating from type I seesaw corresponds to real values of Dirac Yukawa couplings in the terms $Y_{ij} \bar{L_i} H N_j$. This makes the process of right handed heavy neutrino decay into a light neutrino and Higgs $(N \rightarrow \nu H)$ CP preserving ruling out the possibility of leptogenesis. Here we consider the type II seesaw term as the common origin of non-zero $\theta_{13}$ and $\delta_{CP}$ by taking it as a perturbation to the leading order TBM type neutrino mass matrix. First, we numerically fit the type I seesaw term by taking oscillation as well as cosmology data and then compute the predictions for neutrino parameters after the type II seesaw term is introduced. We consider a minimal structure of the type II seesaw term and check whether the predictions for neutrino parameters lie in the $3\sigma$ range. We also compute the predictions for baryon asymmetry of the Universe by considering type II seesaw term as the only source of CP violation and compare it with the latest cosmology data.
\end{abstract}
\maketitle
\section{Introduction}
Several neutrino oscillation experiments in the past few years have provided ample amount of evidence in favor of non-zero yet tiny neutrino masses \cite{PDG}. The smallness (at least ten order of magnitudes compared to the charged fermion masses) of three Standard Model
neutrino masses can be naturally explained 
via seesaw mechanisms which broadly fall into three types : type I \cite{ti}, type II \cite{tii} and type III \cite{tiii}. All these mechanisms involve the inclusion of additional fermionic or scalar fields to generate tiny neutrino masses at tree level. Recent neutrino oscillation experiments T2K \cite{T2K}, Double ChooZ \cite{chooz}, Daya-Bay \cite{daya} and RENO \cite{reno} have not only made the earlier predictions for neutrino parameters more precise, but also predicted non-zero value of the reactor mixing angle $\theta_{13}$. The latest global fit value for $3\sigma$ range of neutrino oscillation parameters \cite{schwetz12} are as follows:
$$ \Delta m_{21}^2=(7.00-8.09) \times 10^{-5} \; \text{eV}^2$$
$$ \Delta m_{31}^2 \;(\text{NH}) =(2.27-2.69)\times 10^{-3} \; \text{eV}^2 $$
$$ \Delta m_{23}^2 \;(\text{IH}) =(2.24-2.65)\times 10^{-3} \; \text{eV}^2 $$
$$ \text{sin}^2\theta_{12}=0.27-0.34 $$
$$ \text{sin}^2\theta_{23}=0.34-0.67 $$ 
\begin{equation}
\text{sin}^2\theta_{13}=0.016-0.030
\end{equation}
where NH and IH refer to normal and inverted hierarchies respectively. The best fit value of leptonic Dirac CP phase $\delta_{CP}$ turns out to be $300$ degrees \cite{schwetz12}.

The neutrino oscillation data before the discovery of non-zero $\theta_{13}$ were in perfect agreement with the so called TBM form of the neutrino mixing matrix widely studied in the literature\cite{Harrison} given by
\begin{equation}
U_{TBM}==\left(\begin{array}{ccc}\sqrt{\frac{2}{3}}&\frac{1}{\sqrt{3}}&0\\
 -\frac{1}{\sqrt{6}}&\frac{1}{\sqrt{3}}&\frac{1}{\sqrt{2}}\\
\frac{1}{\sqrt{6}}&-\frac{1}{\sqrt{3}}& \frac{1}{\sqrt{2}}\end{array}\right),
\end{equation}
which predicts $\text{sin}^2\theta_{12}=\frac{1}{3}$, $\text{sin}^2\theta_{23}=\frac{1}{2}$ and $\text{sin}^2\theta_{13}=0$. However, in view of the fact that the latest experimental data have ruled out $\text{sin}^2\theta_{13}=0$, one needs to go beyond the TBM framework. Since the experimental value of $\theta_{13}$ is still much smaller than the other two mixing angles, TBM can still be a valid approximation and the non-zero $\theta_{13}$ can be accounted for by incorporating the presence of small perturbations to the TBM coming from different sources like charged lepton mass diagonalization, for example. Several such scenarios have been widely discussed in \cite{nzt13, nzt13GA} and the latest neutrino oscillation data can be successfully predicted within the framework of many interesting flavor symmetry models. It will be very interesting if these frameworks which predict non-zero $\theta_{13}$ can also shed some light on the Dirac CP violating phase which is still unknown (and could have remained unknown if $\theta_{13}$ were exactly zero).

Apart from predicting the correct neutrino oscillation data as well as the Dirac CP phase, the nature of neutrino mass hierarchy is also an important yet unresolved issue. Understanding the correct nature of hierarchy can also have non-trivial relevance in leptogenesis, neutrino-less double beta decay experiments among others. The observed baryon asymmetry in the Universe is encoded in the baryon to photon ratio measured by dedicated cosmology experiments like Wilkinson Mass Anisotropy Probe (WMAP), Planck etc. The latest data available from Planck mission constrain the baryon to photon ratio \cite{Planck13} as
\begin{equation}
Y_B \simeq (6.065 \pm 0.090) \times 10^{-10}
\label{barasym}
\end{equation} 
Leptogenesis is one of the most widely studied mechanism of generating this observed baryon asymmetry in the Universe by generating an asymmetry in the leptonic sector first and later converting it into baryon asymmetry through electroweak sphaleron transitions \cite{sphaleron}. As pointed out first by Fukugita and Yanagida \cite{fukuyana}, the out of equilibrium CP violating decay of heavy Majorana neutrinos provides a natural way to create the required lepton asymmetry. The novel feature of this mechanism is the way it relates two of the most widely studied problems in particle physics: the origin of neutrino mass and the origin of matter-antimatter asymmetry. This idea has been implemented in several interesting models in the literature \cite{leptoreview,joshipura,davidsonPR}. Recently such a comparative study was done to understand the impact of mass hierarchies, Dirac and Majorana CP phases on the predictions for baryon asymmetry in \cite{leptodborah} within the framework of left-right symmetric models.

In view of above, the present work is planned to study the possibility to have a common origin of three important observables in neutrino sector: reactor mixing angle $\theta_{13}$, Dirac CP phase $\delta_{CP}$ and baryon asymmetry within the framework of type I and type II seesaw mechanisms. Type I seesaw term gives rise to TBM type neutrino mixing whereas type II seesaw term acts like a perturbation to TBM mixing. We allow this perturbation to be complex such that it can simultaneously generate both $\theta_{13}$ and $\delta_{CP}$. Similar works were recently done in \cite{db-t2} where type II seesaw was considered to be the origin of $\theta_{13}$. Similar attempts to study the deviations from TBM mixing by using the interplay of two different seesaw mechanisms were done in \cite{devtbmt2}. Here we extend our earlier work further to explain Dirac CP phase and leptogenesis together with non-zero $\theta_{13}$. Another work was done recently in \cite{mkd-db-rm} 
where either type I or type II term was considered as leading order and the impact of the other term as a small perturbation on neutrino parameters was studied. In another work \cite{dbgrav}, the impact of Planck suppressed operators on neutrino mixing parameters was studied. In the present work, we assume the leading contribution to neutrino mass (the type I seesaw term) as TBM type which is numerically fitted with the oscillation data on mass squared differences and cosmological upper bound on the sum of absolute neutrino masses. The motivation behind this assumption is the dynamical origin of TBM mixing pattern in terms of a broken flavor symmetry based on discrete groups like $A_4$ \cite{A4TBM}. The type II seesaw term is then introduced as a complex perturbation and the predictions for the neutrino parameters, baryon asymmetry as well as observables like sum of absolute neutrino masses $\sum_i \lvert m_i \rvert$, effective neutrino mass $m_{ee} = \lvert \sum_i U^2_{ei} m_i \rvert$ etc. are calculated. We vary the strength of this perturbation and check whether the same strength of the perturbation can generate non-zero $\theta_{13}$ in agreement with experiments and also keep the other neutrino parameters as well as baryon asymmetry within the allowed range. We also check whether the sum of absolute neutrino masses obey the cosmological upper bound $\sum_i \lvert m_i \rvert < 0.23$ eV \cite{Planck13} as we vary the strength of the perturbation. We consider both normal and inverted hierarchical neutrino mass patterns as well as two different values of the lightest active neutrino mass eigenstate. By estimating the required type II seesaw strength so as to produce sufficient deviation from TBM mixing we constrain the mass of the additional Higgs triplet $M_{\Delta}$ responsible for type II seesaw. The constraint turns out to be very close to the grand unification scale $M_{\Delta} \sim 10^{16}$ GeV.

This paper is organized as follows: in section \ref{method} we discuss the methodology of type I and type II seesaw mechanisms. In section \ref{devTBM}, we discuss the parametrization of TBM type $\mu-\tau$ symmetric neutrino mass matrix as well as the deviations from TBM mixing in order to generate non-zero reactor mixing angle. In section \ref{sec:lepto}, we outline the mechanism of leptogenesis in the presence of type I and type II seesaw. In section \ref{numeric} we discuss our numerical analysis and results and then finally conclude in section \ref{conclude}.
\section{Seesaw Mechanism: Type I and Type II}
\label{method}
Type I seesaw \cite{ti} mechanism is the simplest possible realization of the dimension five Weinberg operator \cite{weinberg} for the origin of neutrino masses in a renormalizable theory. This mechanism is implemented in the standard model by the inclusion of three additional right handed neutrinos $(\nu^i_R, i = 1,2,3)$ as $SU(2)_L$ singlets with zero $U(1)_Y$ charges. Being singlet under the gauge group, bare mass terms of the right handed neutrinos $M_{RR}$ are allowed in the Lagrangian. The resulting type I seesaw formula for light neutrinos is given by the expression,
\begin{equation}
m_{LL}^I=-m_{LR}M_{RR}^{-1}m_{LR}^{T}.
\end{equation}
where $m_{LR}$ is the Dirac mass term of the neutrinos which is typically of electroweak scale. Demanding the light neutrinos to be of eV scale one needs $M_{RR}$ to be as high as $10^{14}$ GeV without any fine-tuning of Dirac Yukawa couplings. 

On the other hand, in type II seesaw \cite{tii} mechanism, the standard model is extended by inclusion of an additional $SU(2)_L$ triplet scalar field having $U(1)_Y$ charge twice that of lepton doublets. It can be represented as
\begin{equation}
\Delta_L =
\left(\begin{array}{cc}
\ \delta^+_L/\surd 2 & \delta^{++}_L \\
\ \delta^0_L & -\delta^+_L/\surd 2
\end{array}\right) \nonumber
\end{equation} 
The gauge charges of this field allow a new term in the Yukawa Lagrangian
$ f_{ij}\ \left(\ell_{iL}^T \ C \ i \sigma_2 \Delta_L \ell_{jL}\right)$ which can account for tiny neutrino masses if the neutral component of the scalar triplet $\delta^0_L$ acquires a tiny vacuum expectation value (vev). From the minimization of the scalar potential, it turns out that the vev of $\delta^0_L$ is given by 
\begin{equation}
 \langle \delta^0_L \rangle = v_L = \frac{\mu_{\Delta H}\langle \phi^0 \rangle^2}{M^2_{\Delta}}
\label{vev} 
\end{equation}
where $\phi^0=v$ is the neutral component of the electroweak Higgs doublet with vev approximately $10^2$ GeV. The trilinear coupling term $\mu_{\Delta H}$ and the mass term of the triplet $M_{\Delta}$ can be taken to be of same order. Thus, $M_{\Delta}$ has to be as high as $10^{14}$ GeV to give rise to tiny neutrino masses without any fine-tuning of the dimensionless couplings $f_{ij}$.
\begin{table}
\centering
\caption{Parametrization of the neutrino mass matrix for TBM mixing}
\vspace{0.5cm}
{\small
\begin{tabular}{|c|c|c|c|c|}
 \hline
   Parameters & IH(Case I) &  IH(Case II) &  NH(Case I)&  NH(Case II)\\ \hline
x&0.0487942&0.0812709&0.0035726&0.0701779\\  \hline
y&0.0002555&0.0001536&0.0025726&0.0001778\\  \hline
z&-0.023769&-0.008058&0.0243546&0.0079243\\  \hline
$m_3\;(\text{eV})$&0.001&0.065&0.049&0.085\\  \hline
$m_2\; (\text{eV})$&0.04930&0.08157&0.00871&0.07053\\  \hline
$m_1\;(\text{eV})$&0.04853&0.08111&0.001&0.07\\  \hline
$\sum_i m_i\;(\text{eV})$&0.2225&0.2225&0.2059&0.2059\\  \hline
\end{tabular}
}
\label{table:results1}
\end{table}
\section{Deviations from TBM mixing}
\label{devTBM}
The $\mu-\tau$ symmetric TBM type neutrino mass matrix originating from type I seesaw can be parametrized as
\begin{equation}
m_{LL}=\left(\begin{array}{ccc}
x& y&y\\
y& x+z & y-z \\
y & y-z & x+z
\end{array}\right)
\label{matrix1}
\end{equation}
which is clearly $\mu-\tau$ symmetric with eigenvalues $m_1 = x-y, \; m_2 = x+2y, \; m_3 = x-y+2z$. It predicts the mixing angles as $\theta_{12} \simeq 35.3^o, \; \theta_{23} = 45^o$ and $\theta_{13} = 0$. Although the prediction for first two mixing angles are still allowed from oscillation data, $\theta_{13}=0$ has been ruled out experimentally at more than $9\sigma$ confidence level. This has also opened up the possibility of measuring leptonic CP phase in ongoing as well as future experiments. Non-zero leptonic CP phase will not only provide us with a better understanding of the origin of lepton masses and mixing but could also be responsible for creating baryon asymmetry of the Universe through leptogenesis. Discovery of non-zero $\theta_{13}$ has led to a significant number of interesting works trying to explain its possible origin. Here we study the possibility of explaining the deviations from TBM mixing and hence from $\theta_{13}=0$ by allowing the type II seesaw term as a perturbation. We consider this perturbation to be complex in nature so that it can simultaneously introduce a phase in the neutrino mass matrix which takes the form of $\delta_{CP}$ in the leptonic mixing matrix.

Before choosing the minimal structure of the type II seesaw term, we note that the parametrization of the TBM plus corrected neutrino mass matrix can be done as \cite{nzt13GA}.
\begin{equation}
m_{LL}=\left(\begin{array}{ccc}
x& y-w&y+w\\
y-w& x+z+w & y-z \\
y+w & y-z& x+z-w
\end{array}\right)
\label{matrix2}
\end{equation}
where $w$ denotes the deviation of $m_{LL}$ from that within TBM frameworks and setting it to zero, the above matrix boils down to the familiar $\mu-\tau$ symmetric matrix (\ref{matrix1}). Thus, the minimal structure of the perturbation term to the leading order $\mu-\tau$ symmetric TBM neutrino mass matrix can be taken as
\begin{equation}
m^{II}_{LL}=\left(\begin{array}{ccc}
0& -w& w\\
-w& w & 0 \\
w & 0& -w
\end{array}\right)
\label{matrix3}
\end{equation}
Such a structure of the type II seesaw term can be explained by continuous as well as discrete flavor symmetries as discussed in one of our earlier works (first reference in \cite{db-t2}).

\begin{figure}
\begin{center}
$\begin{array}{cc}
\includegraphics[width=0.5\textwidth]{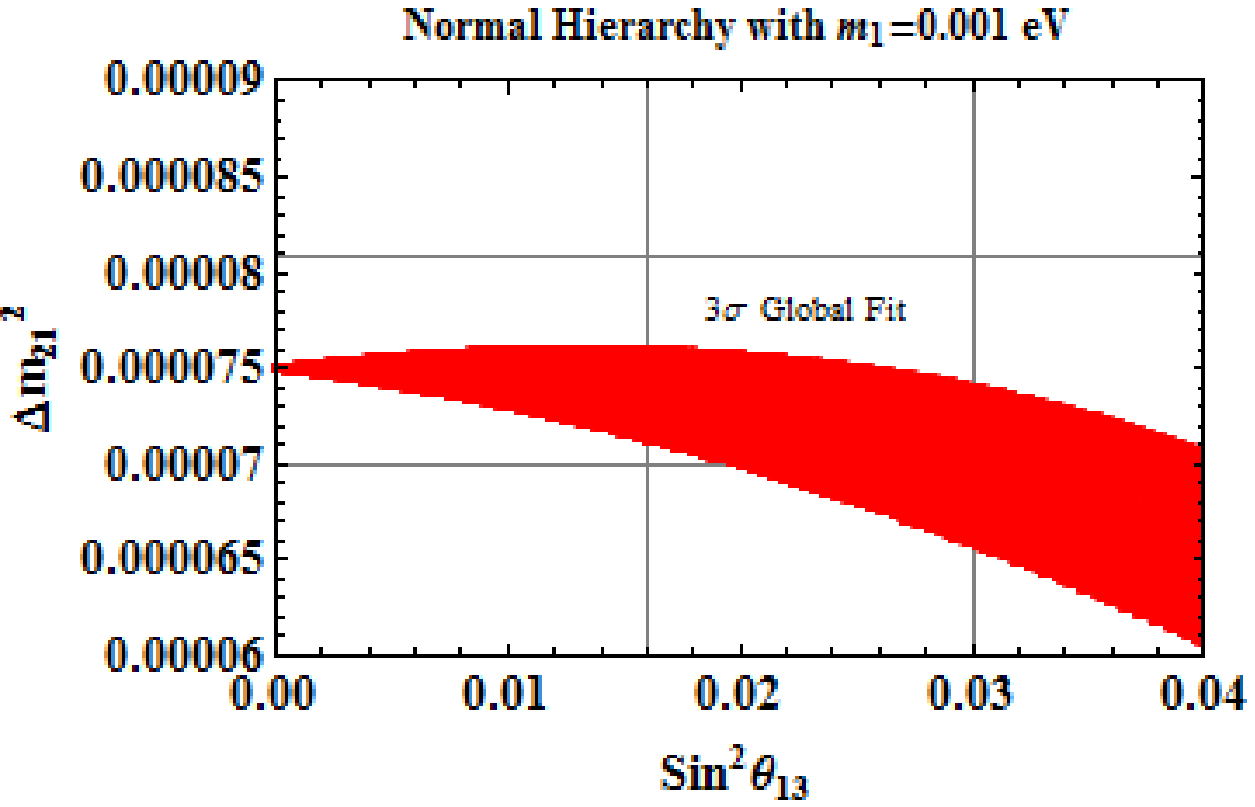} &
\includegraphics[width=0.5\textwidth]{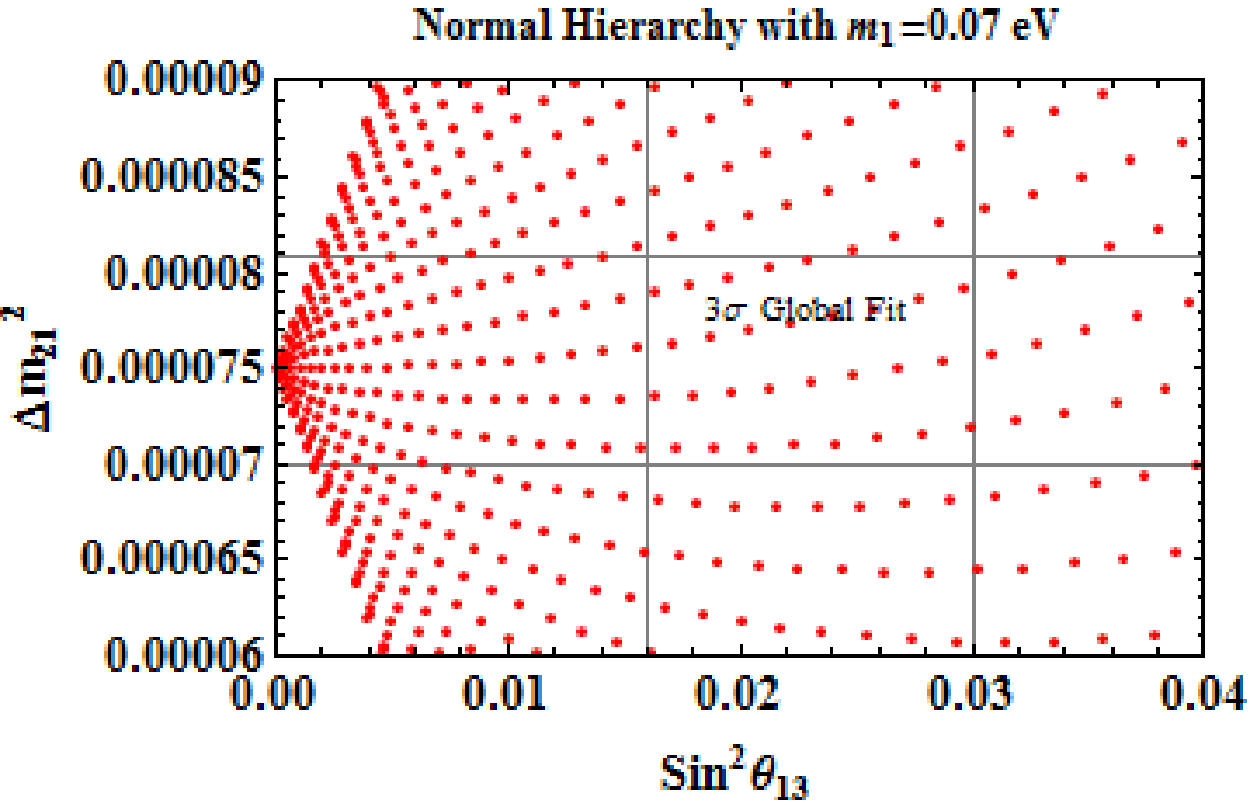} \\
\includegraphics[width=0.5\textwidth]{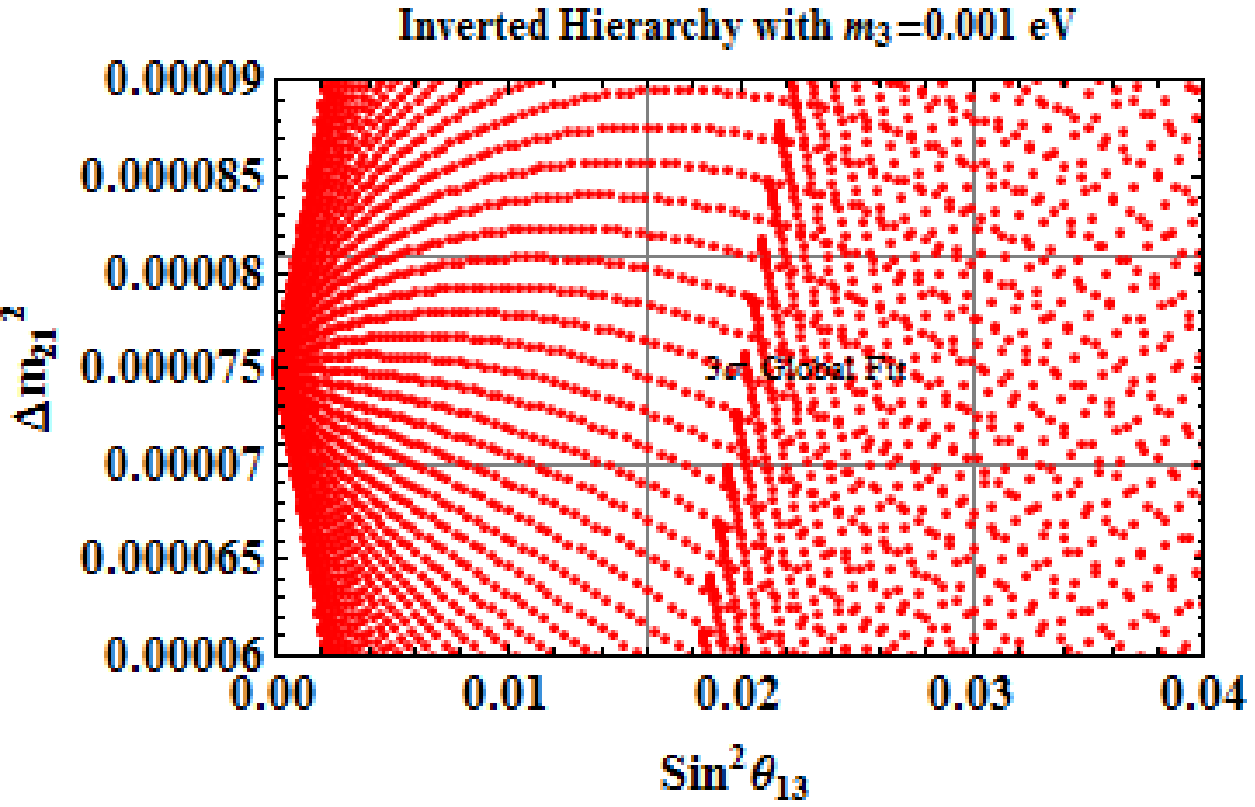} &
\includegraphics[width=0.5\textwidth]{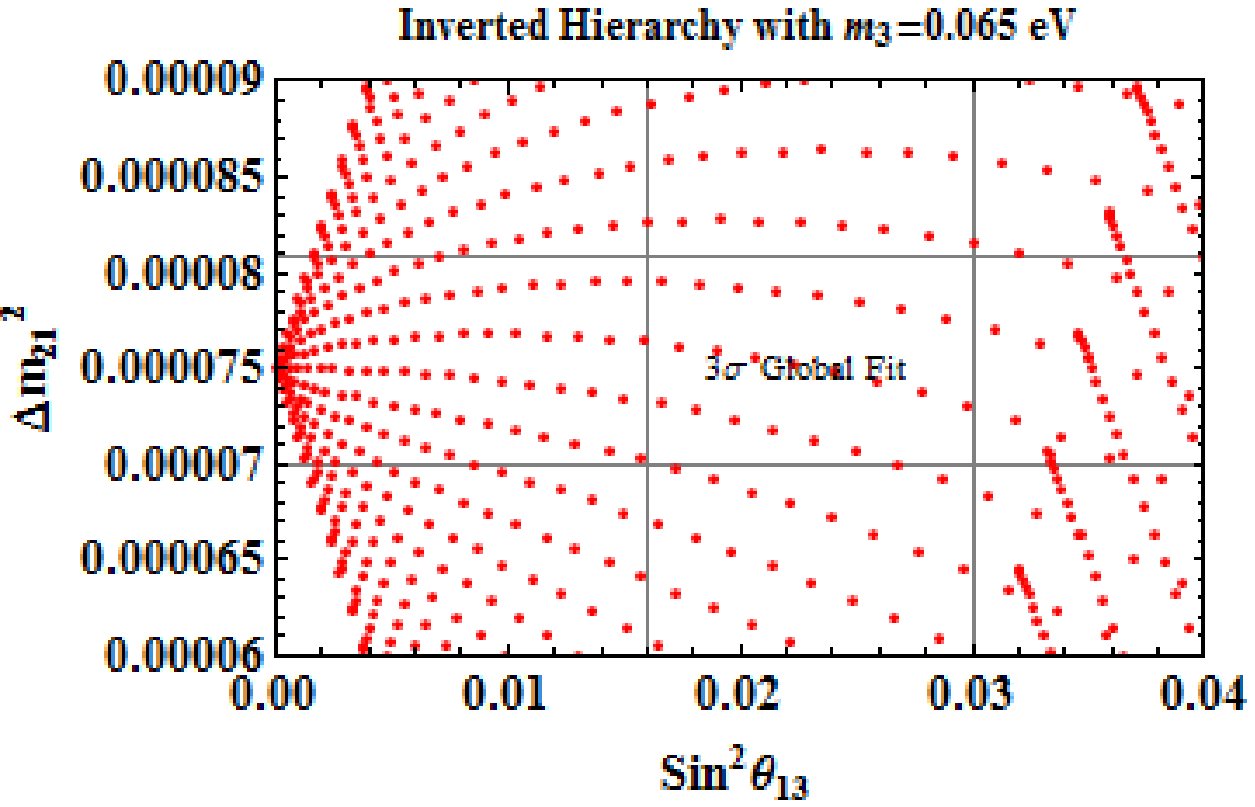}
\end{array}$
\end{center}
\caption{Variation of $\Delta m^2_{21}$ with $\sin^2{\theta_{13}}$}
\label{fig1}
\end{figure}
\begin{figure}[h]
\begin{center}
$
\begin{array}{cc}
\includegraphics[width=0.5\textwidth]{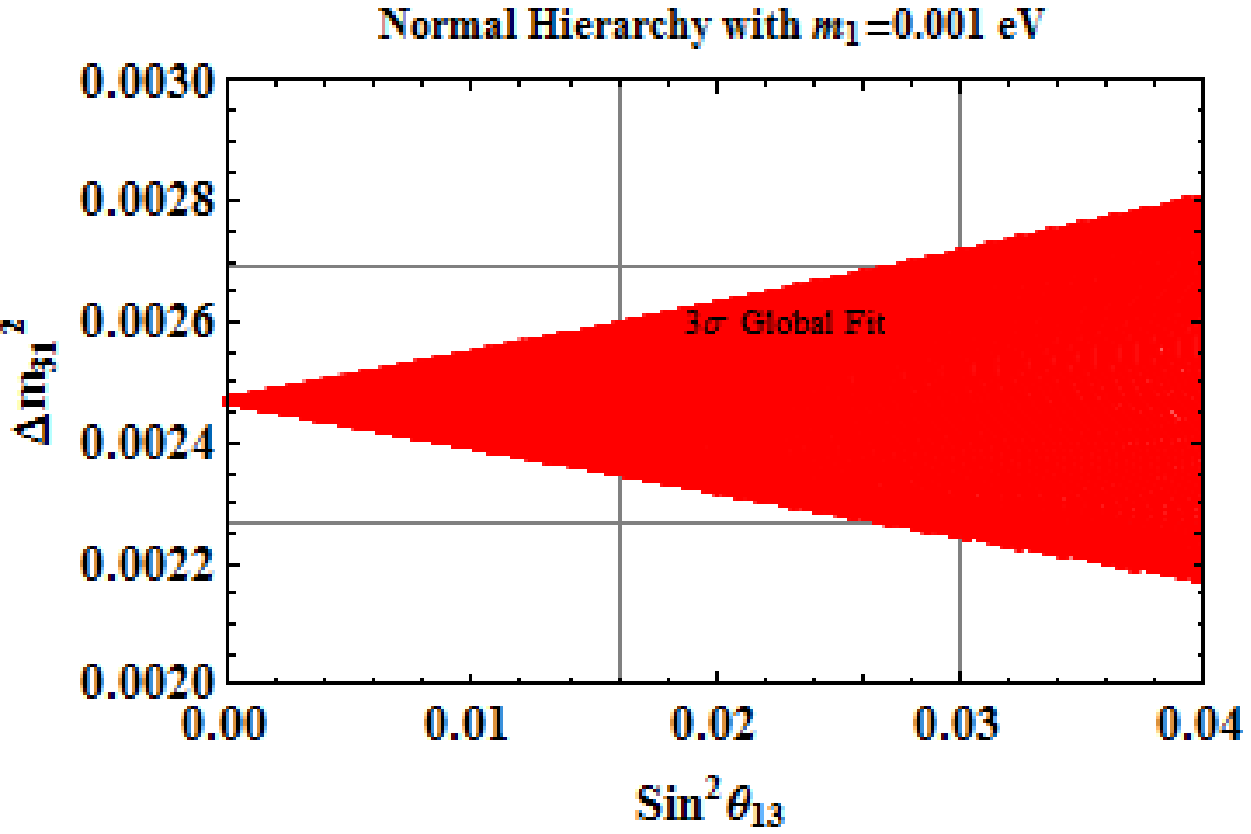} &
\includegraphics[width=0.5\textwidth]{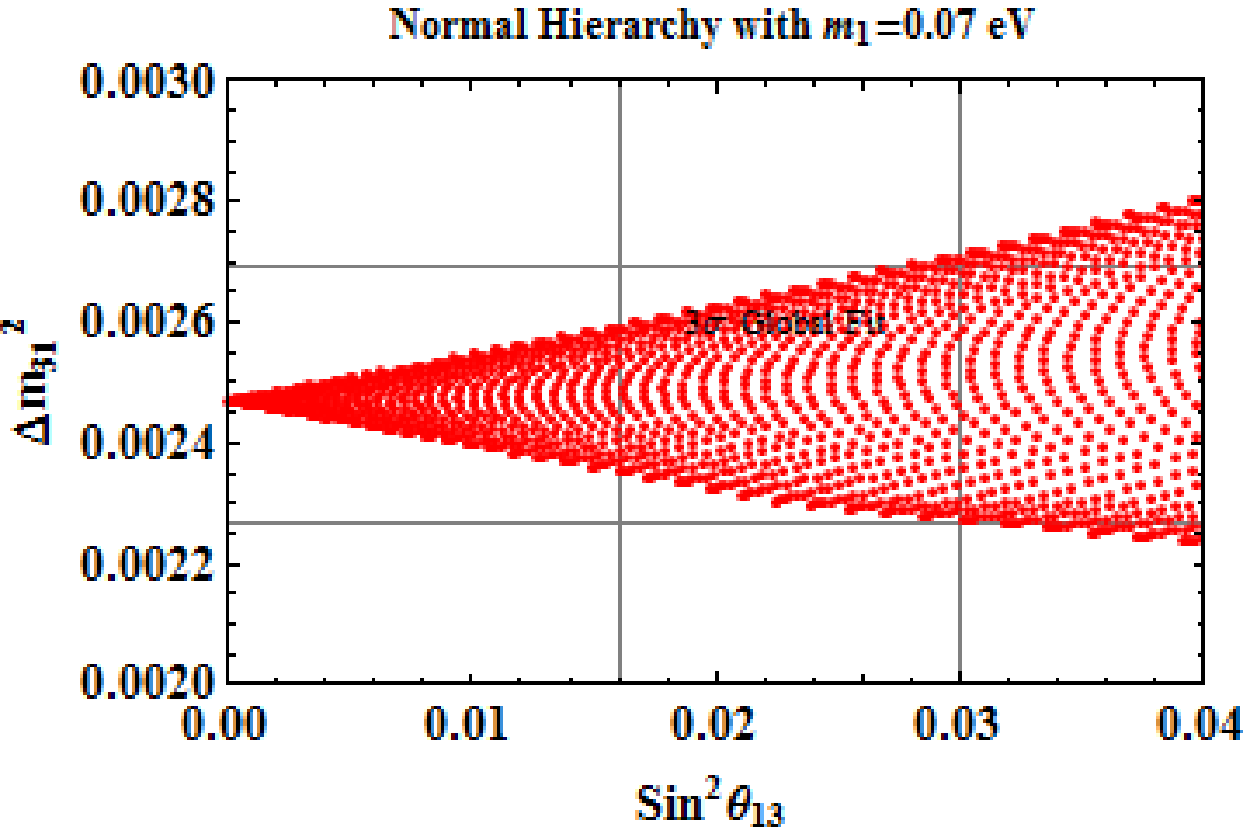} \\
\includegraphics[width=0.5\textwidth]{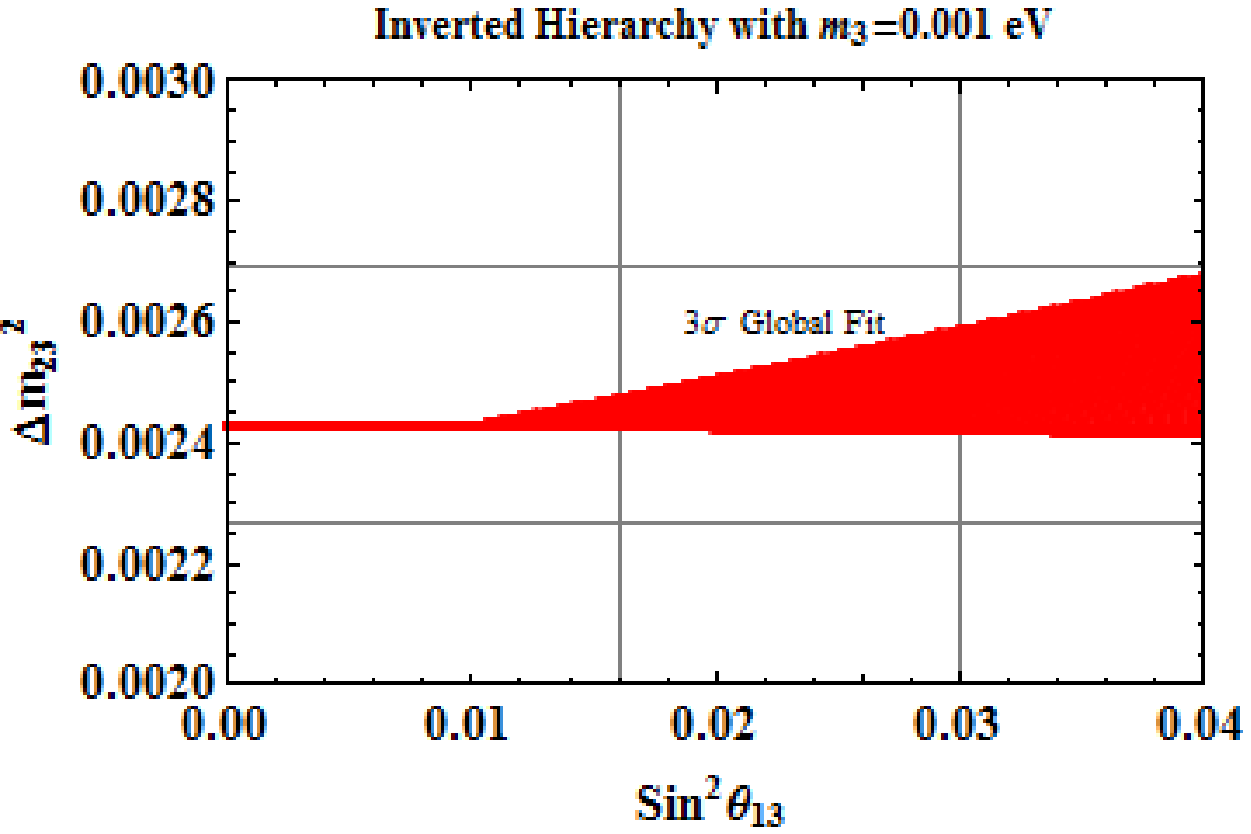} &
\includegraphics[width=0.5\textwidth]{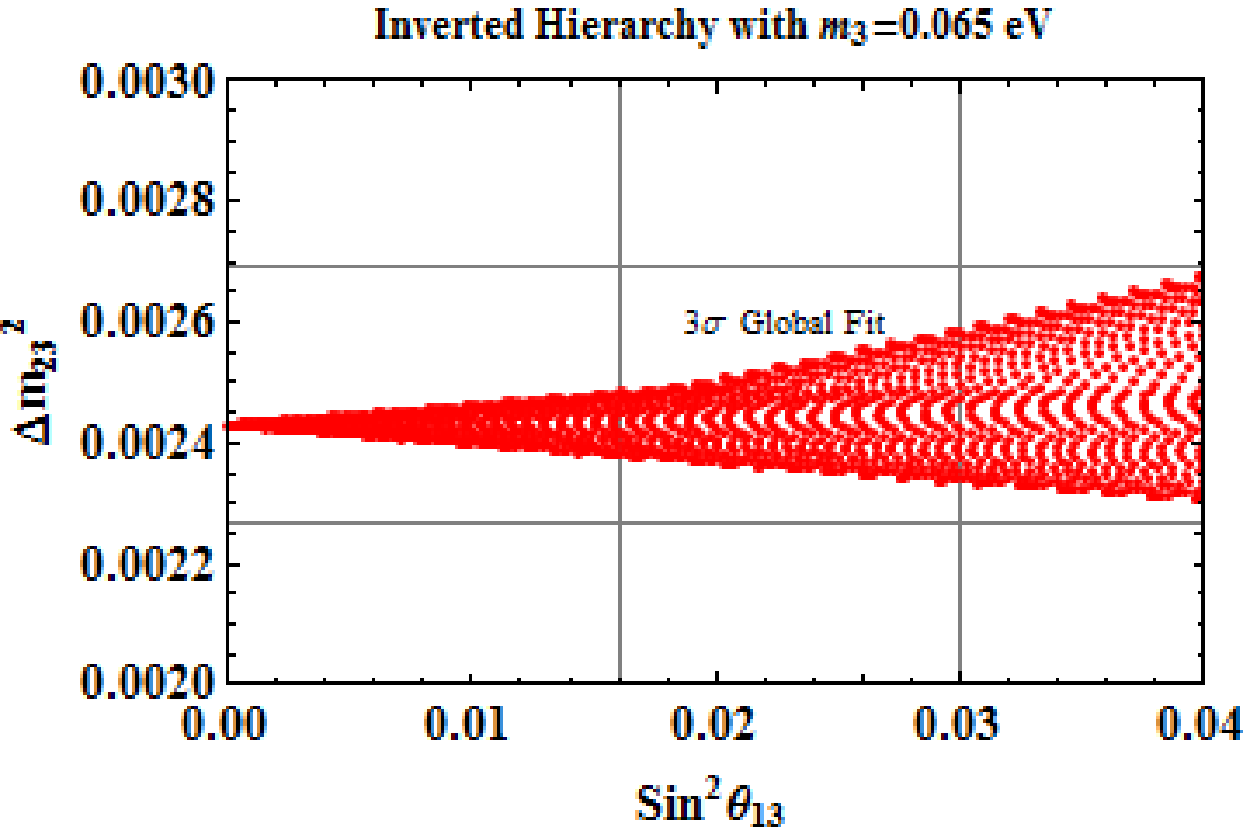}
\end{array}$
\end{center}
\caption{Variation of $\Delta m^2_{31}$ (NH), $\Delta m^2_{23}$ (IH) with $\sin^2{\theta_{13}}$}
\label{fig2}
\end{figure}
\begin{figure}[h]
\begin{center}
$
\begin{array}{cc}
\includegraphics[width=0.5\textwidth]{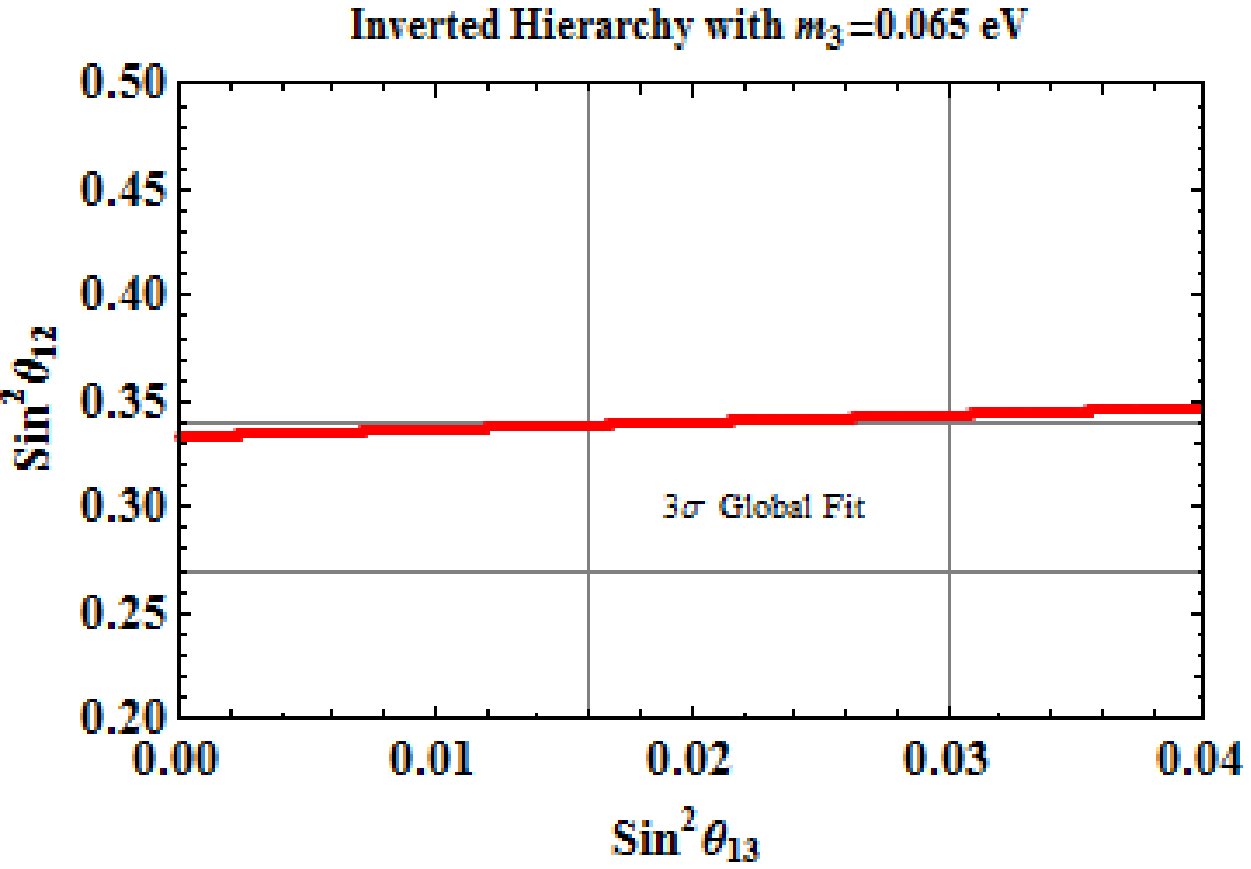} &
\includegraphics[width=0.5\textwidth]{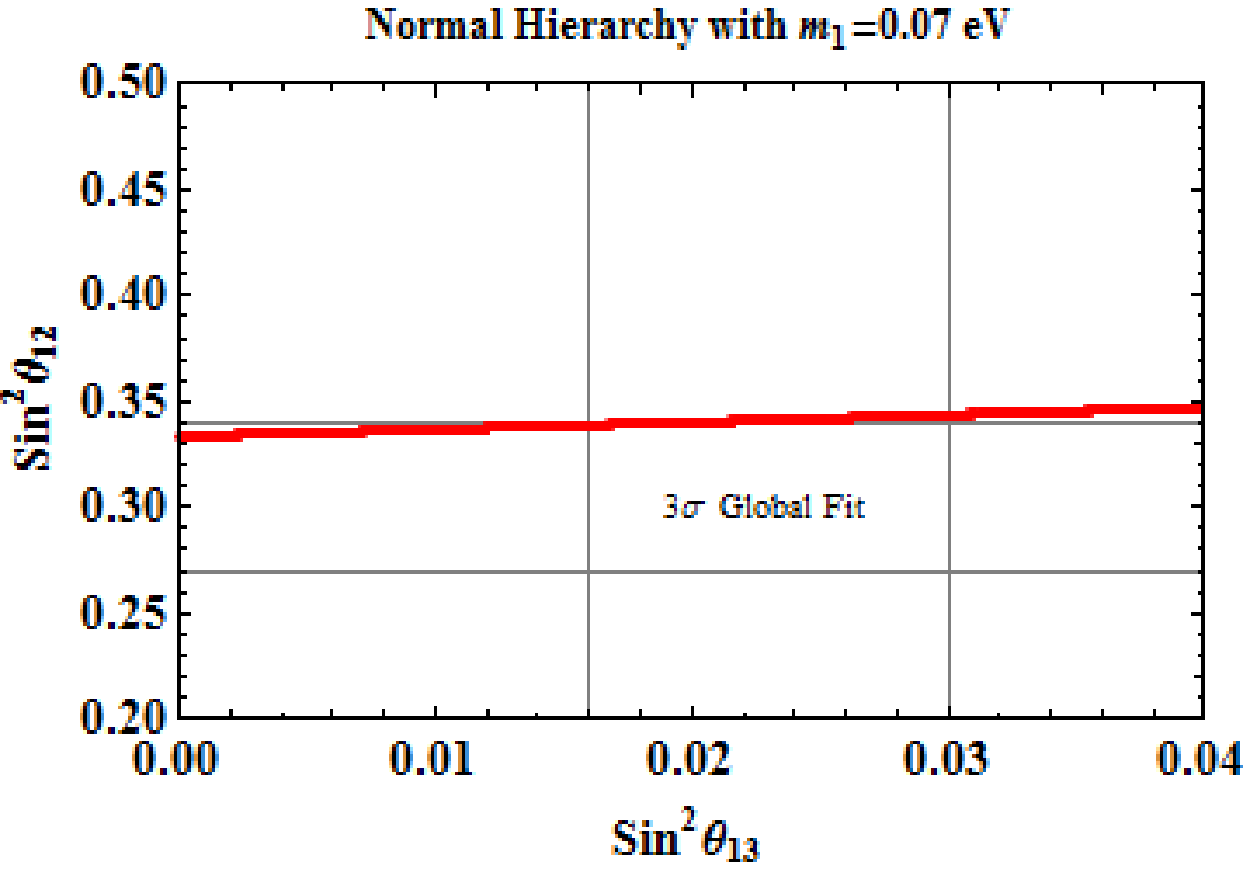} \\
\includegraphics[width=0.5\textwidth]{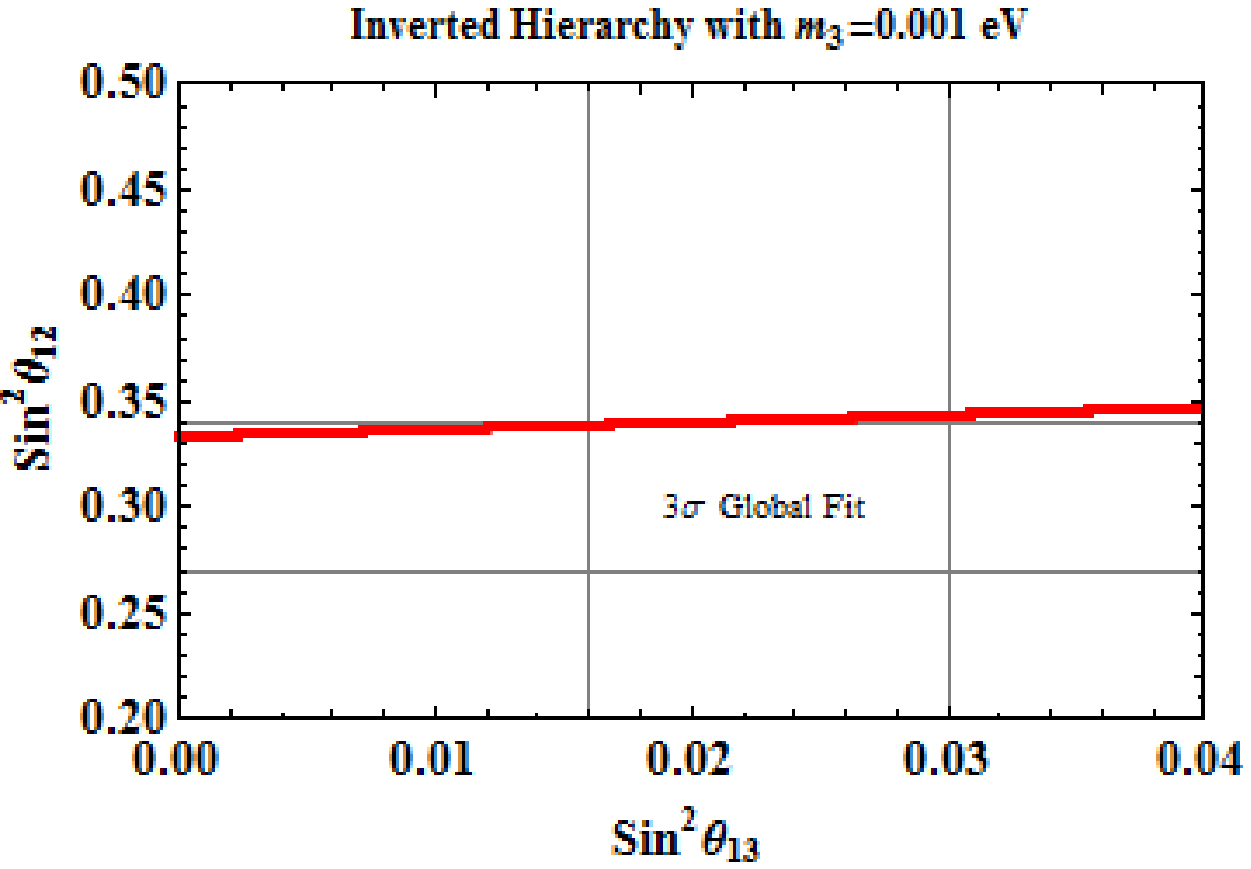} &
\includegraphics[width=0.5\textwidth]{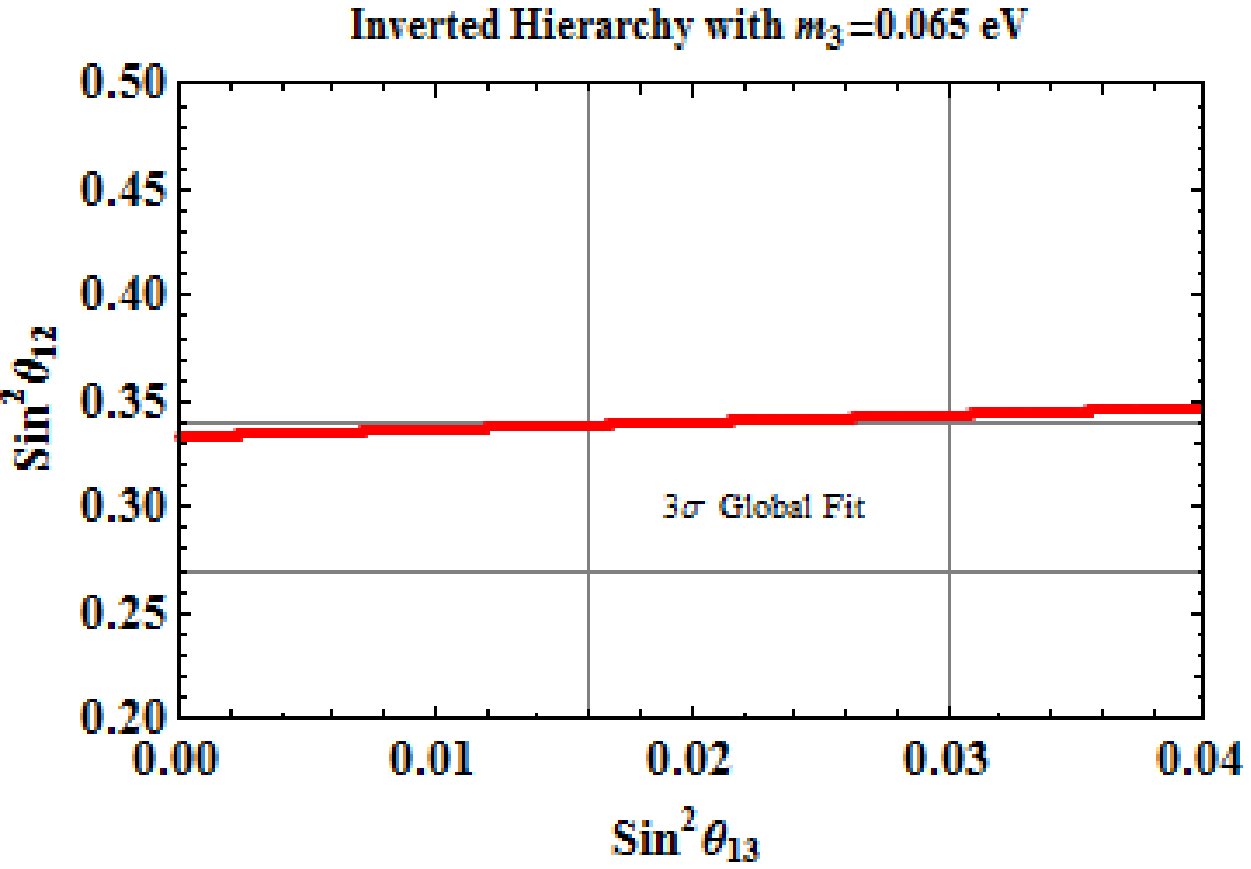}
\end{array}$
\end{center}
\caption{Variation of $\sin^2{\theta_{12}}$ with $\sin^2{\theta_{13}}$}
\label{fig3}
\end{figure}
\begin{figure}[h]
\begin{center}
$
\begin{array}{cc}
\includegraphics[width=0.5\textwidth]{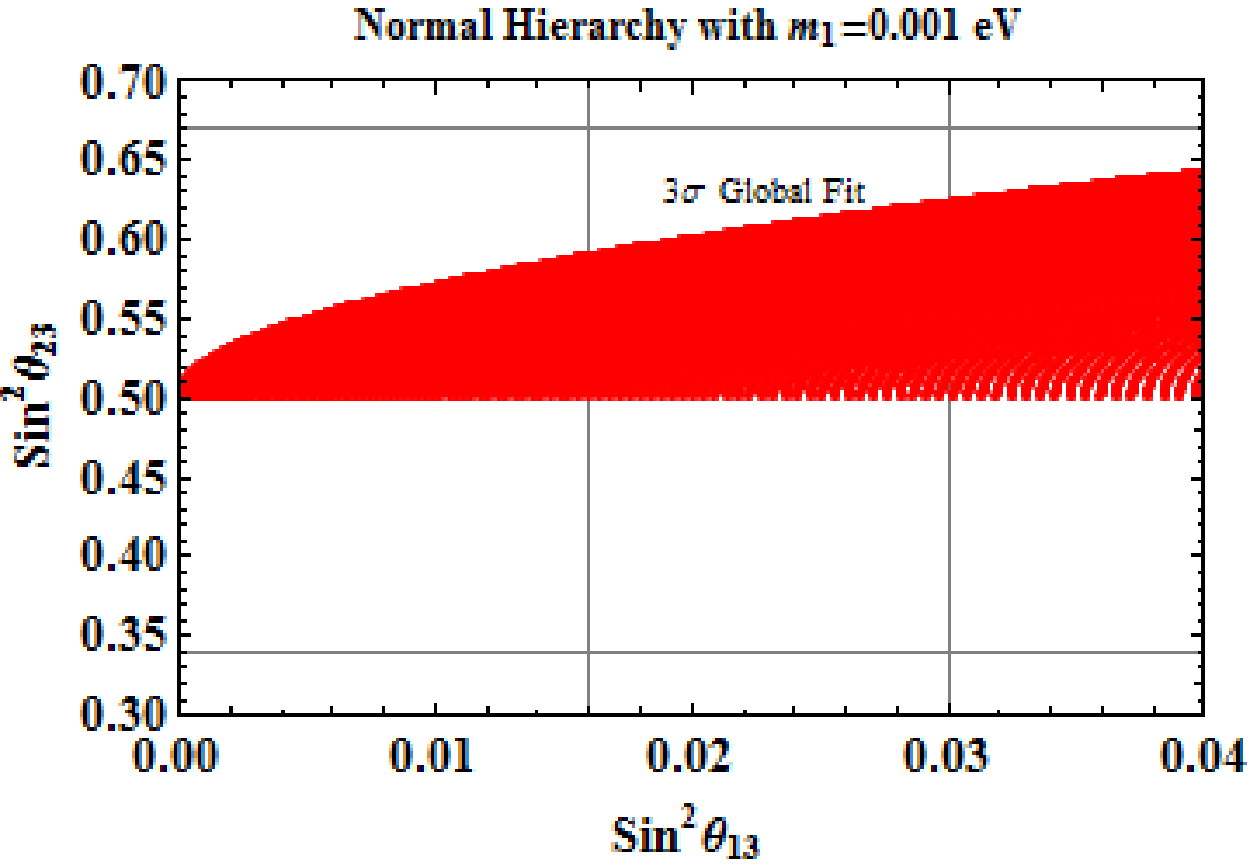} &
\includegraphics[width=0.5\textwidth]{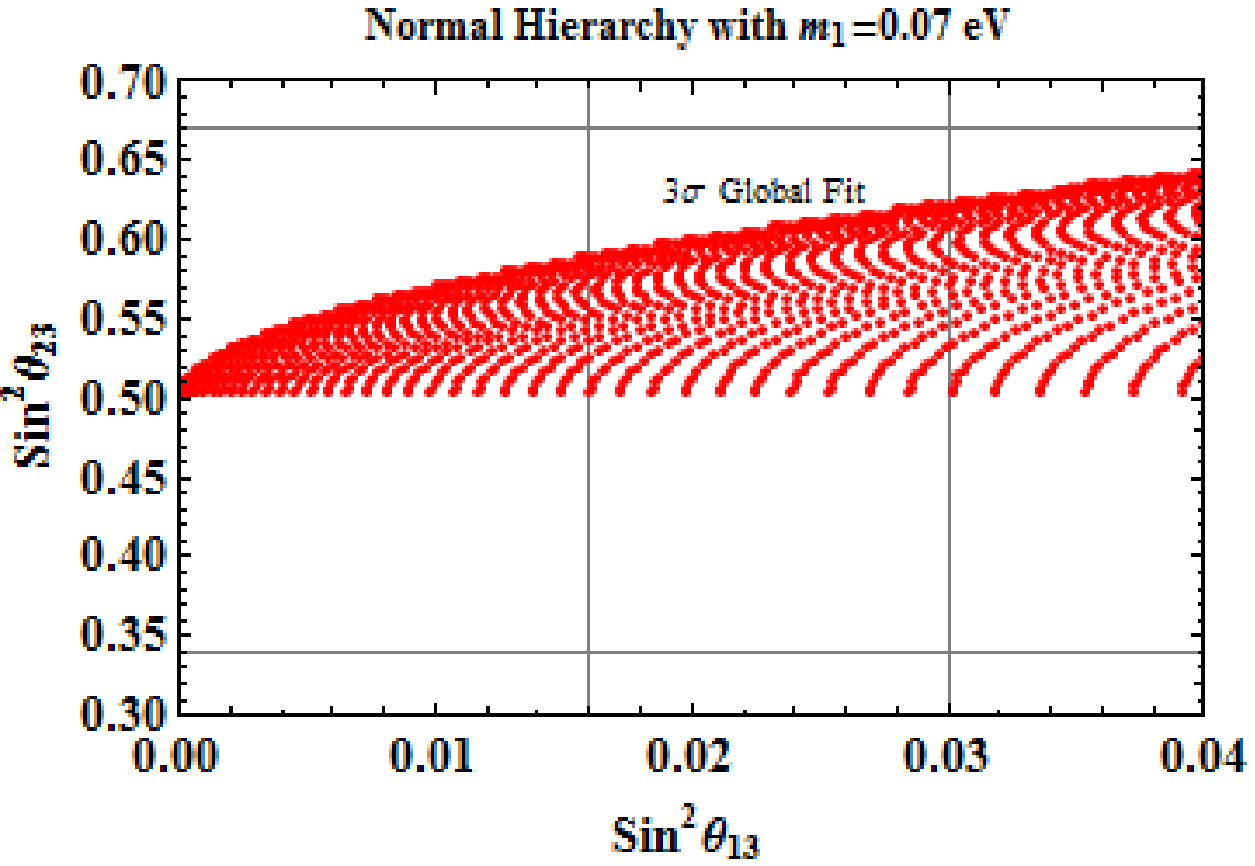} \\
\includegraphics[width=0.5\textwidth]{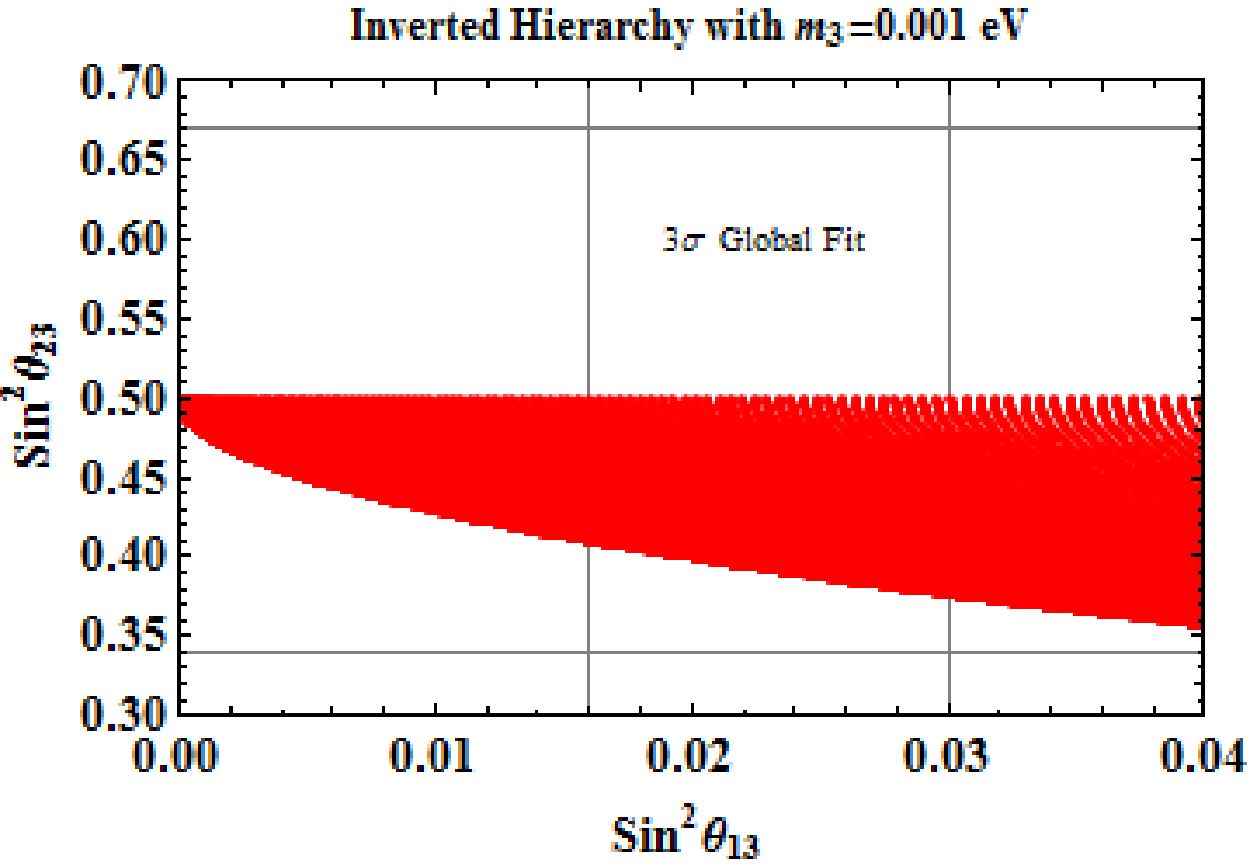} &
\includegraphics[width=0.5\textwidth]{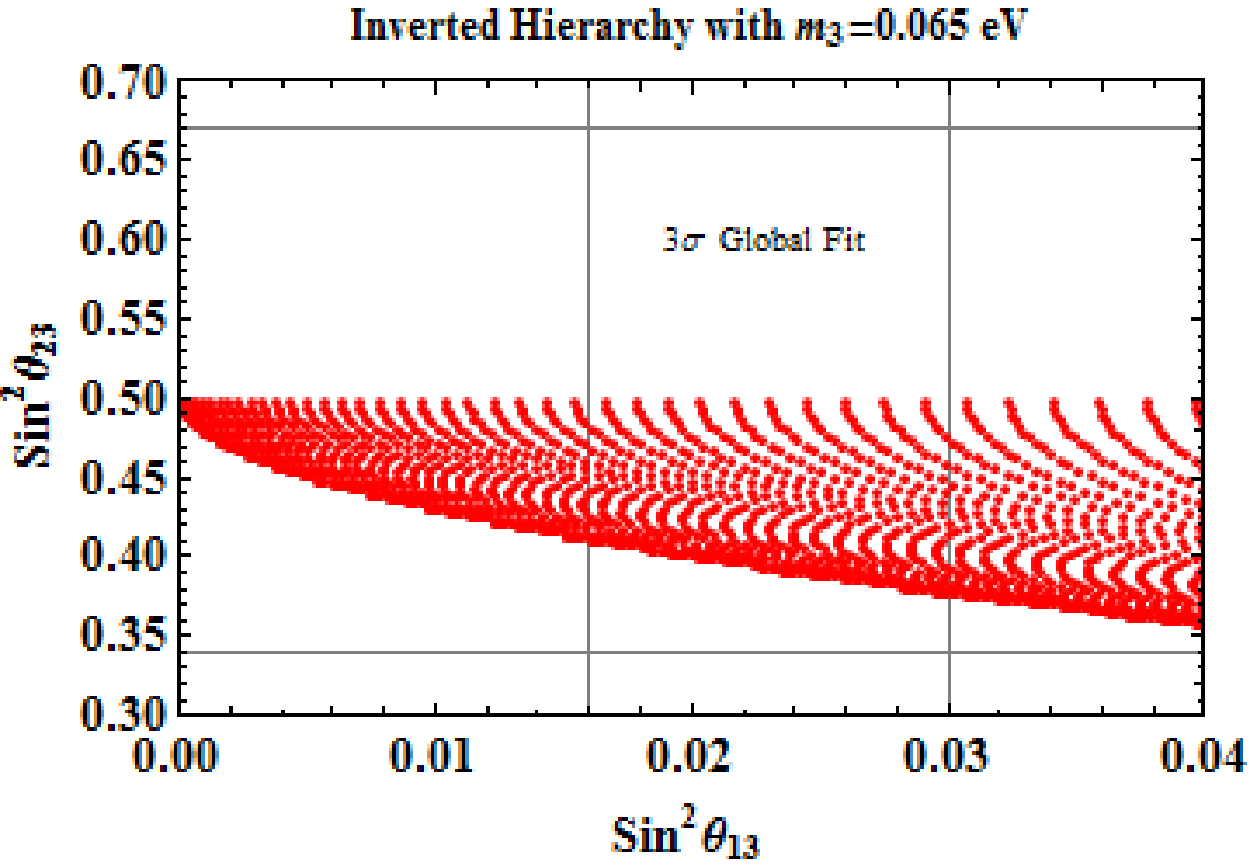}
\end{array}$
\end{center}
\caption{Variation of $\sin^2{\theta_{23}}$ with $\sin^2{\theta_{13}}$}
\label{fig4}
\end{figure}
\begin{figure}[h]
\begin{center}
$
\begin{array}{cc}
\includegraphics[width=0.5\textwidth]{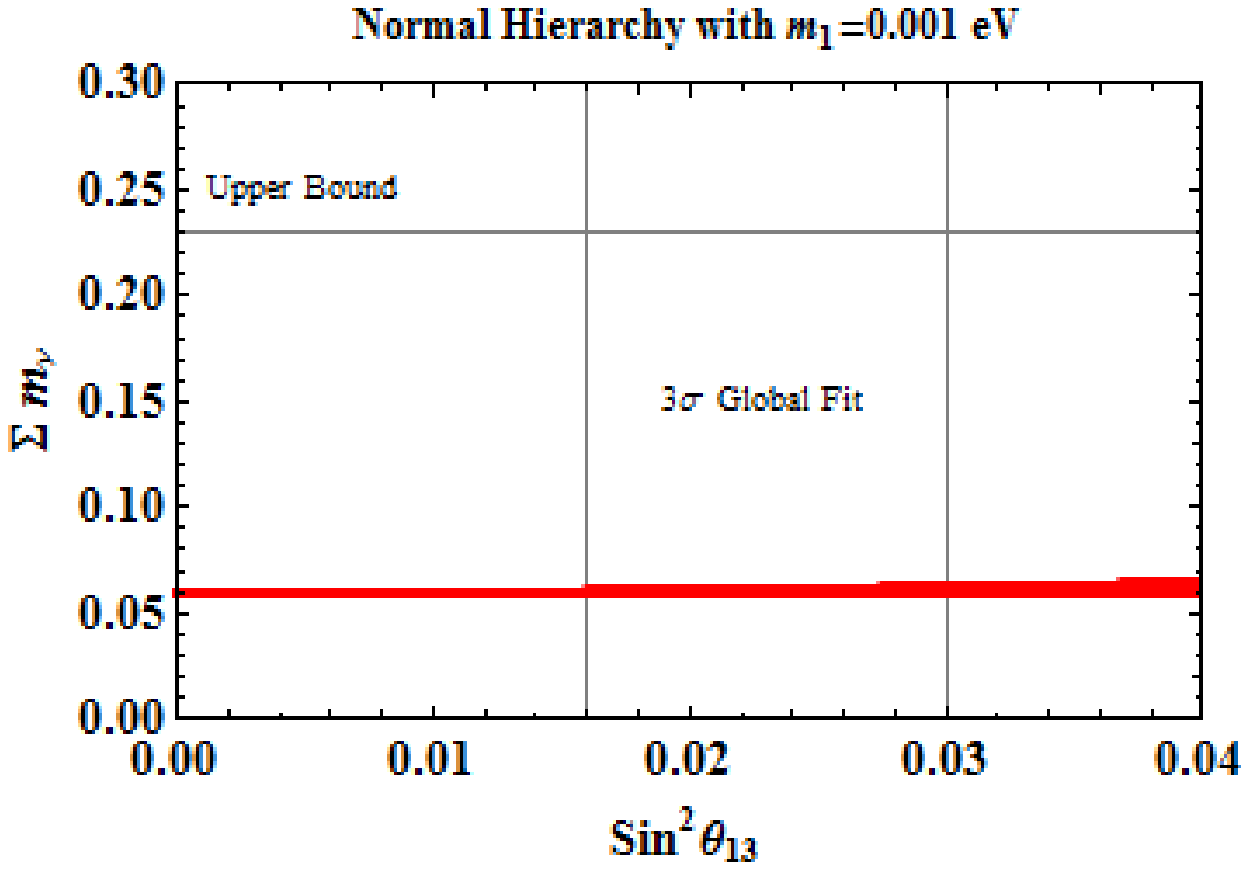} &
\includegraphics[width=0.5\textwidth]{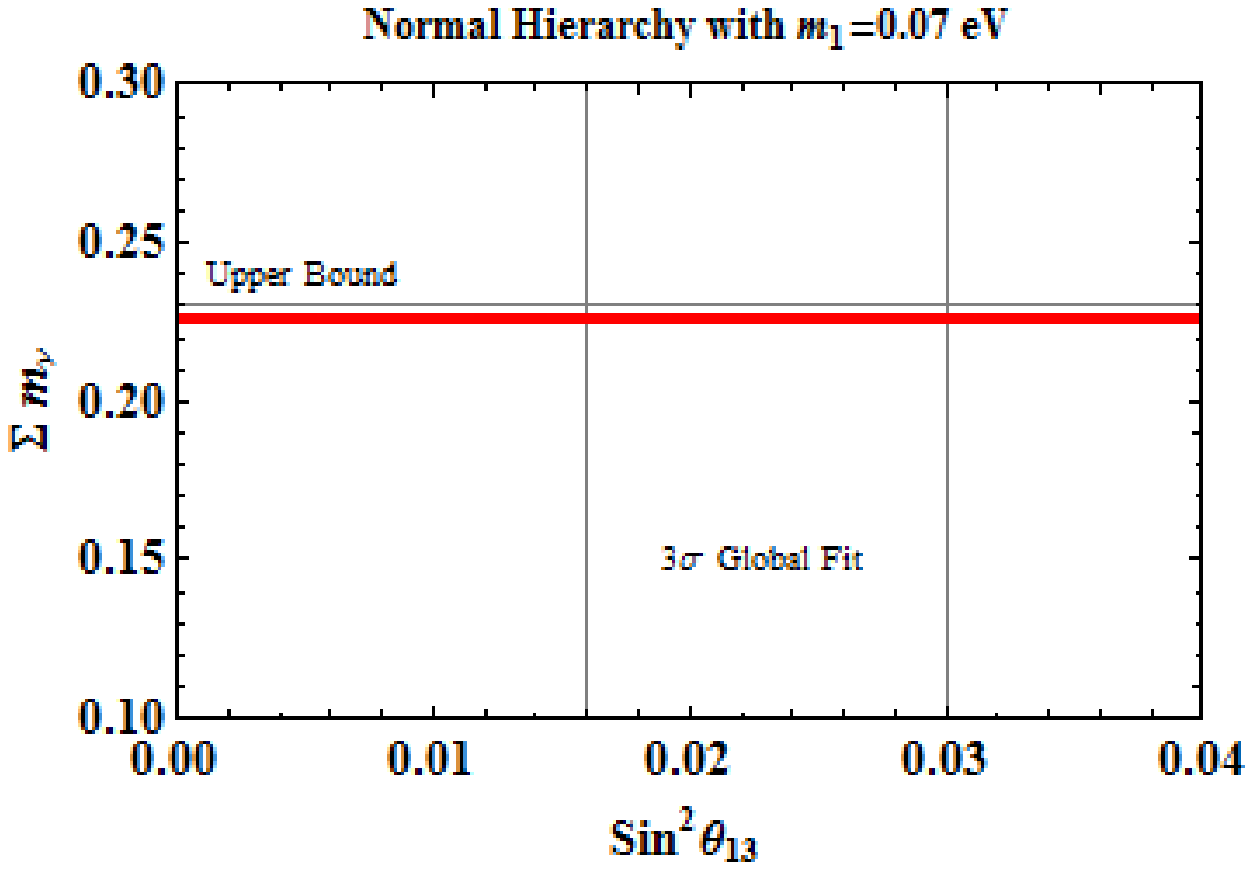} \\
\includegraphics[width=0.5\textwidth]{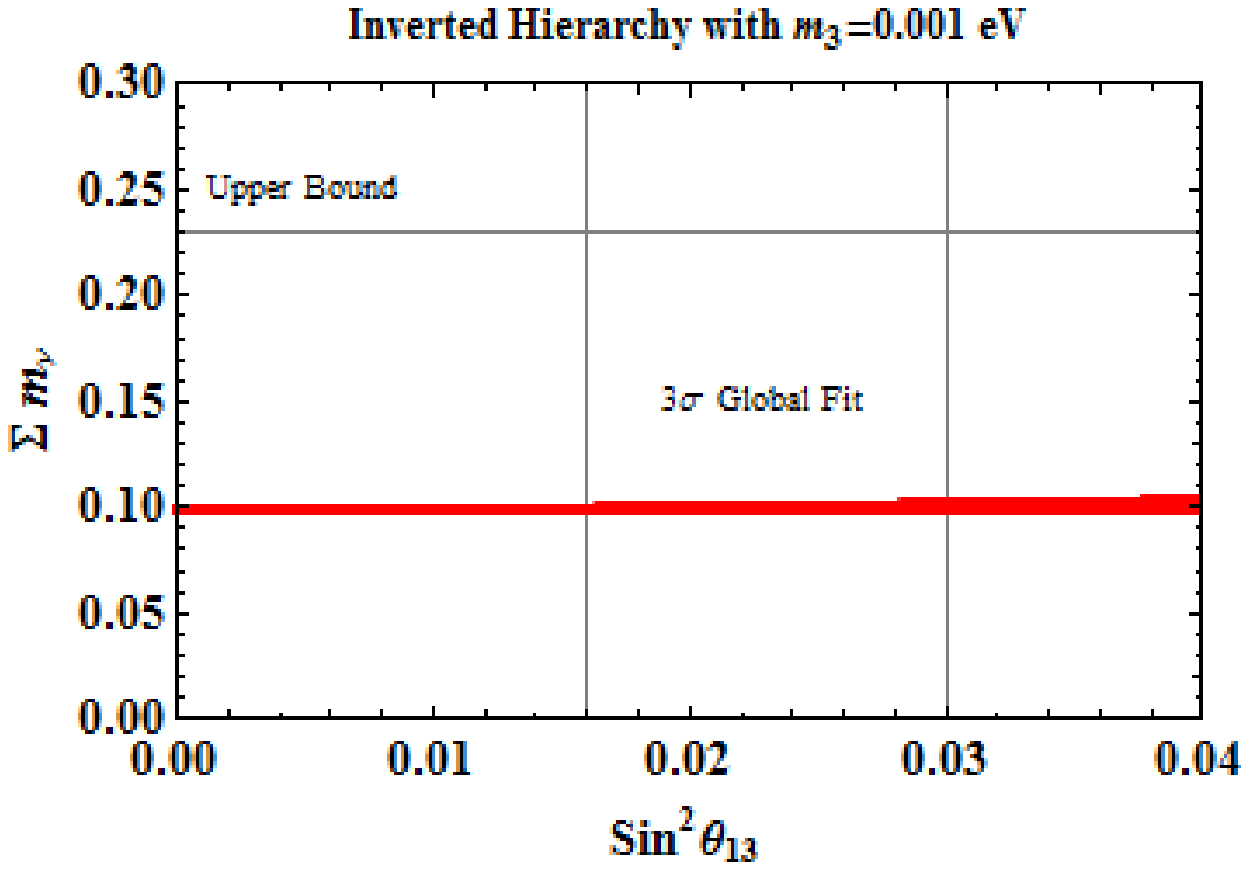} &
\includegraphics[width=0.5\textwidth]{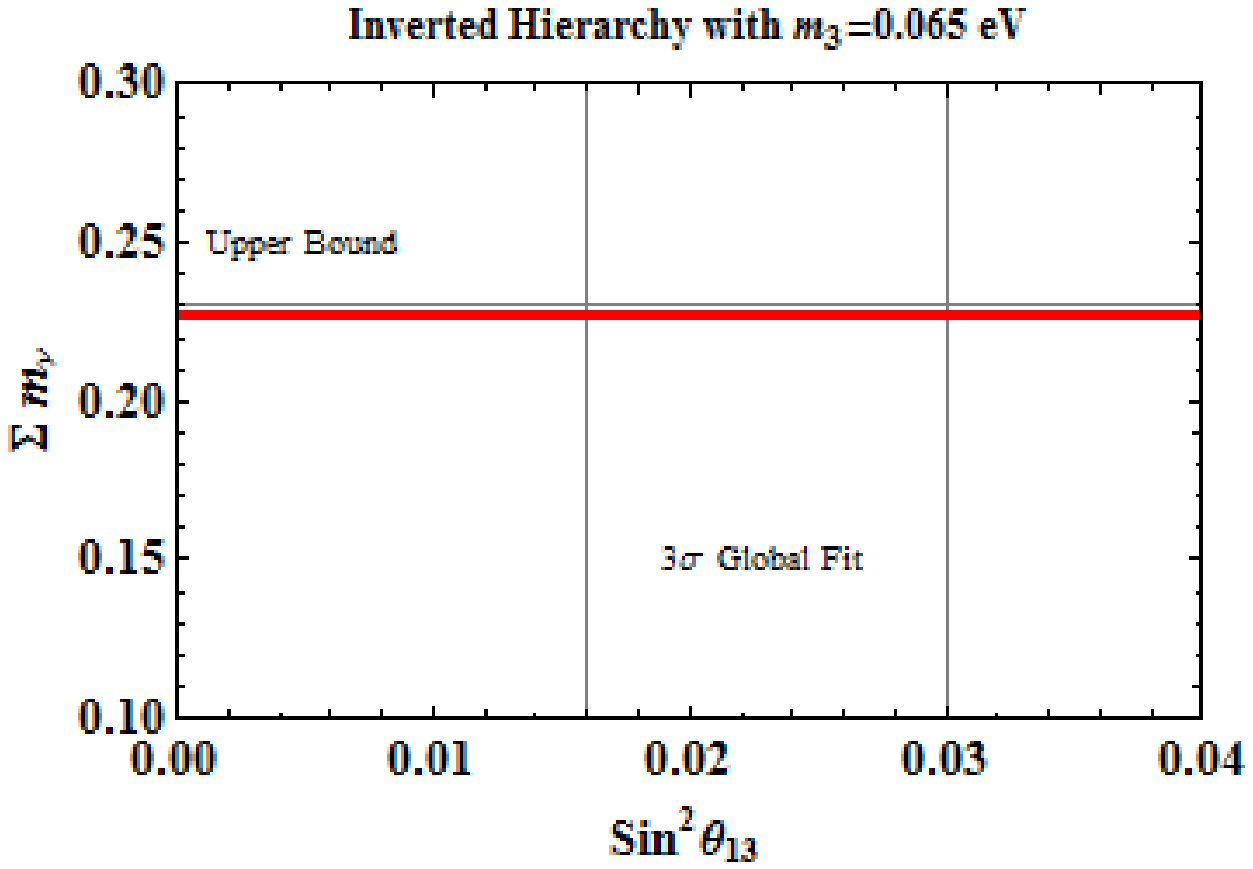}
\end{array}$
\end{center}
\caption{Variation of the sum of absolute neutrino masses with $\sin^2{\theta_{13}}$}
\label{fig5}
\end{figure}
\begin{figure}[h]
\begin{center}
$
\begin{array}{cc}
\includegraphics[width=0.5\textwidth]{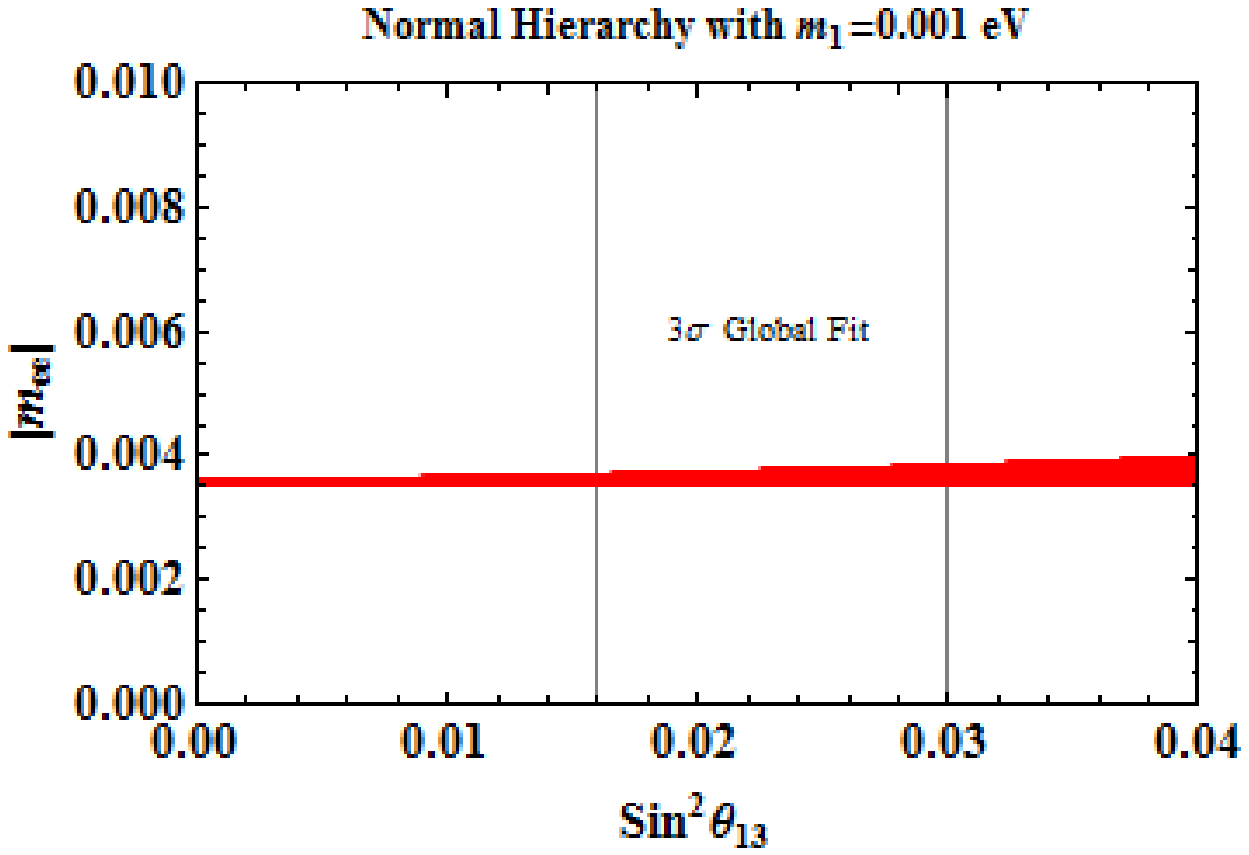} &
\includegraphics[width=0.5\textwidth]{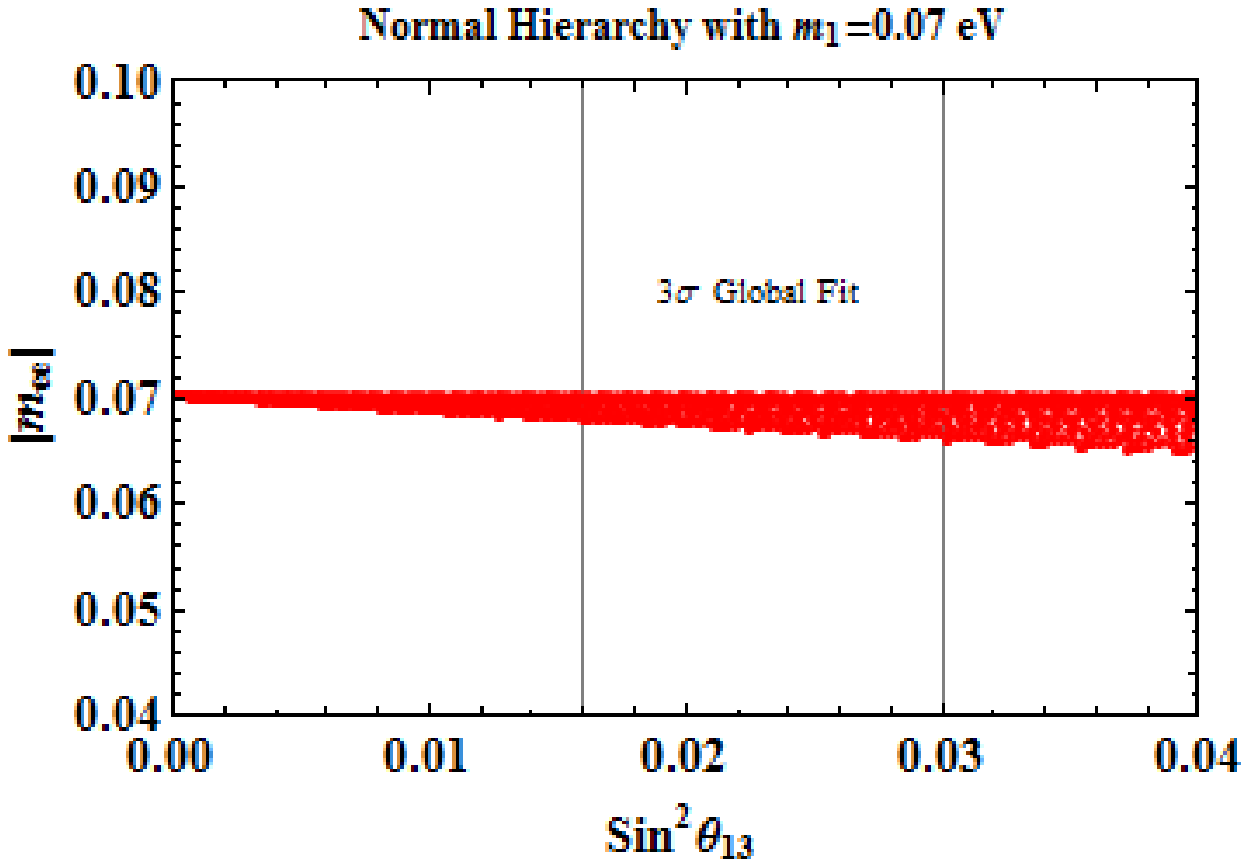} \\
\includegraphics[width=0.5\textwidth]{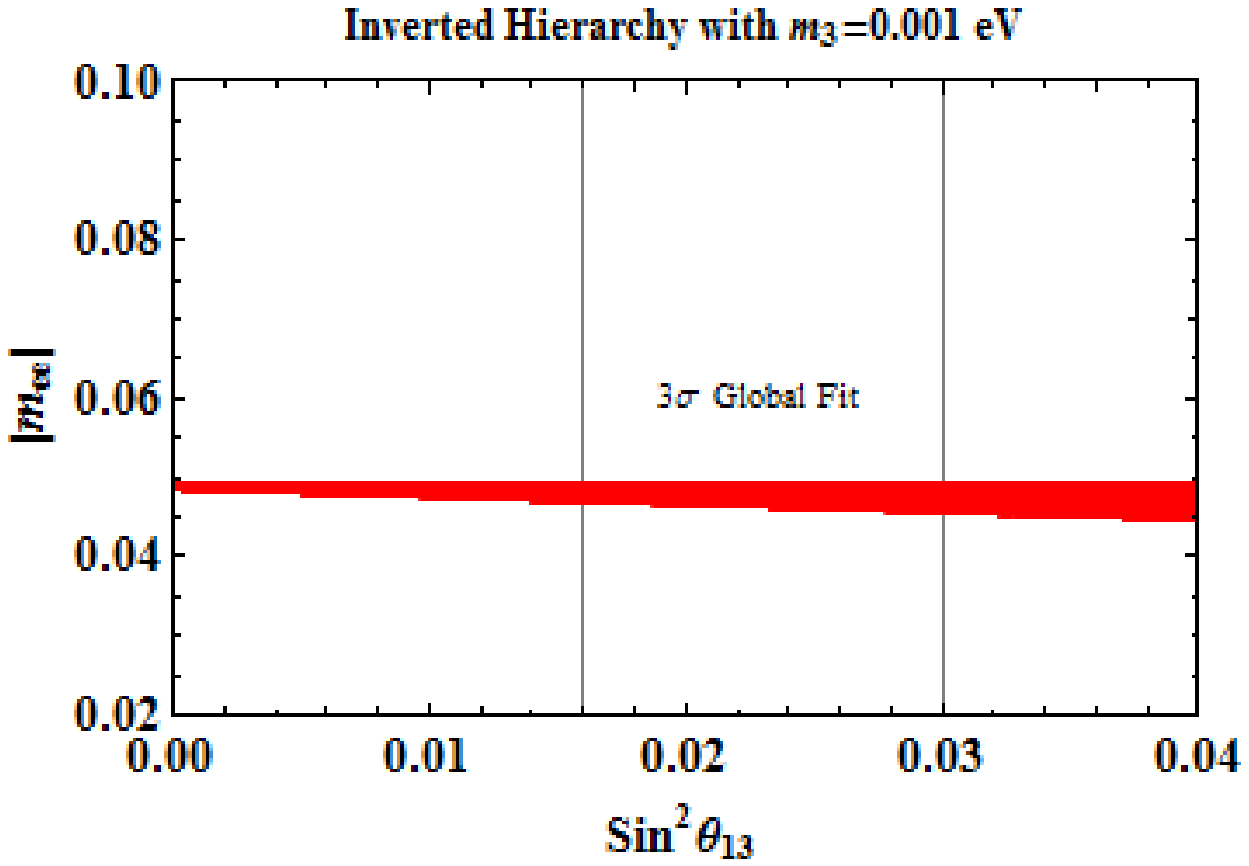} &
\includegraphics[width=0.5\textwidth]{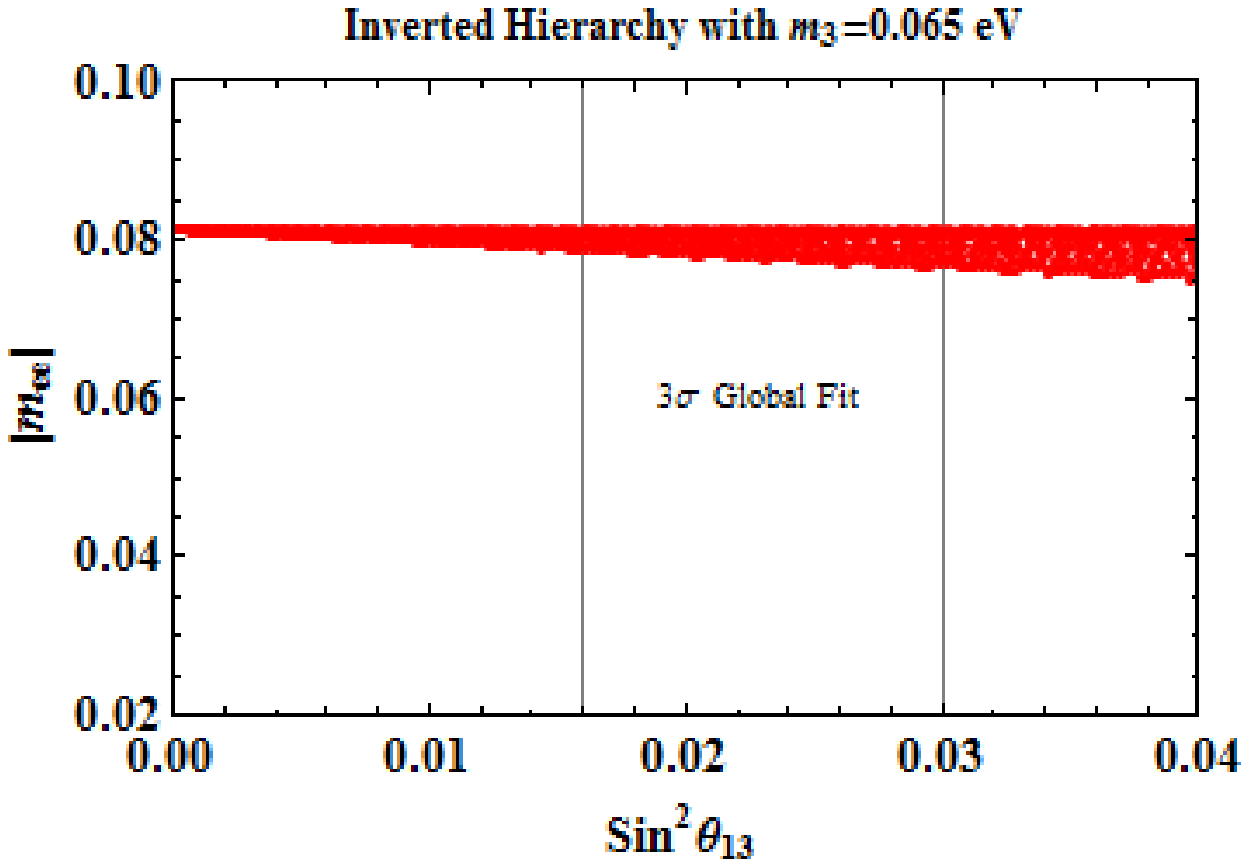}
\end{array}$
\end{center}
\caption{Variation of the effective neutrino mass $m_{ee}$ with $\sin^2{\theta_{13}}$}
\label{fig6}
\end{figure}
\begin{figure}[h]
\begin{center}
$
\begin{array}{cc}
\includegraphics[width=0.5\textwidth]{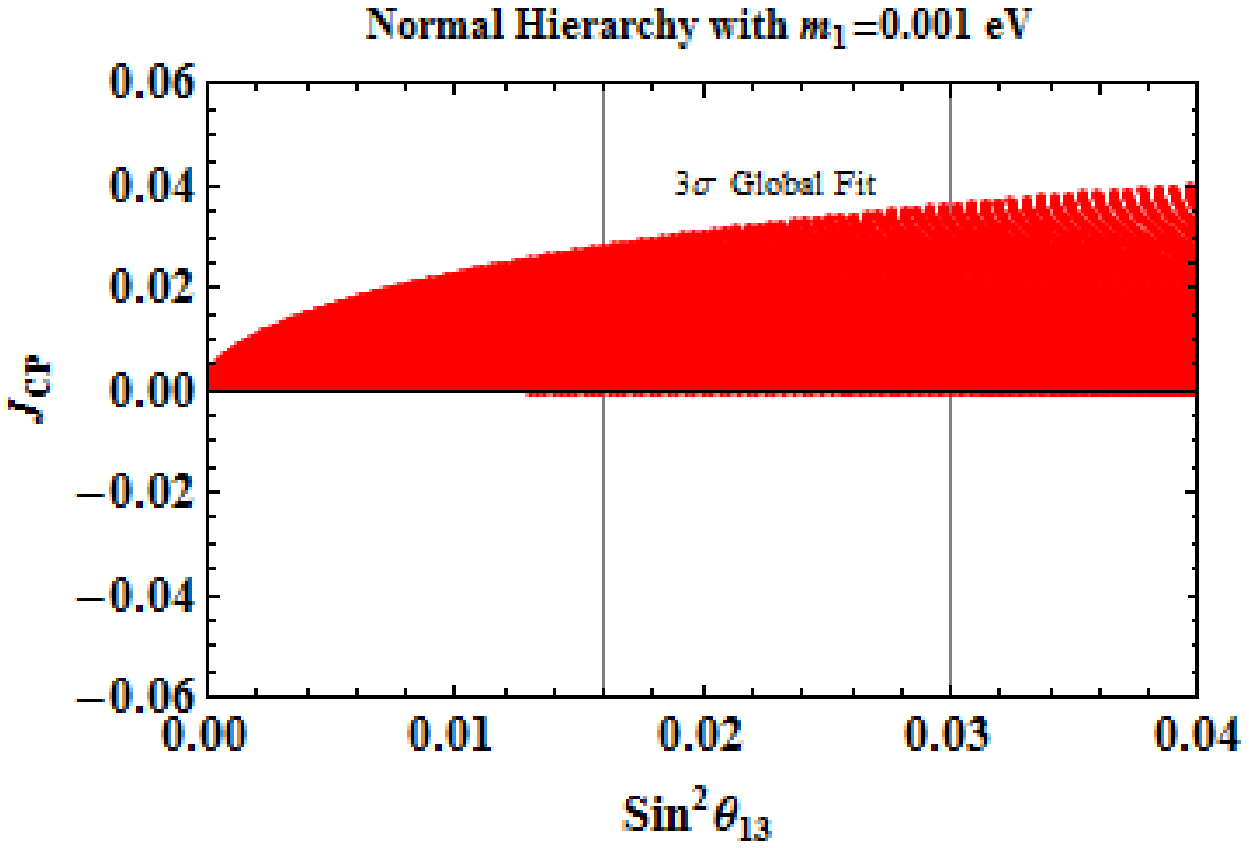} &
\includegraphics[width=0.5\textwidth]{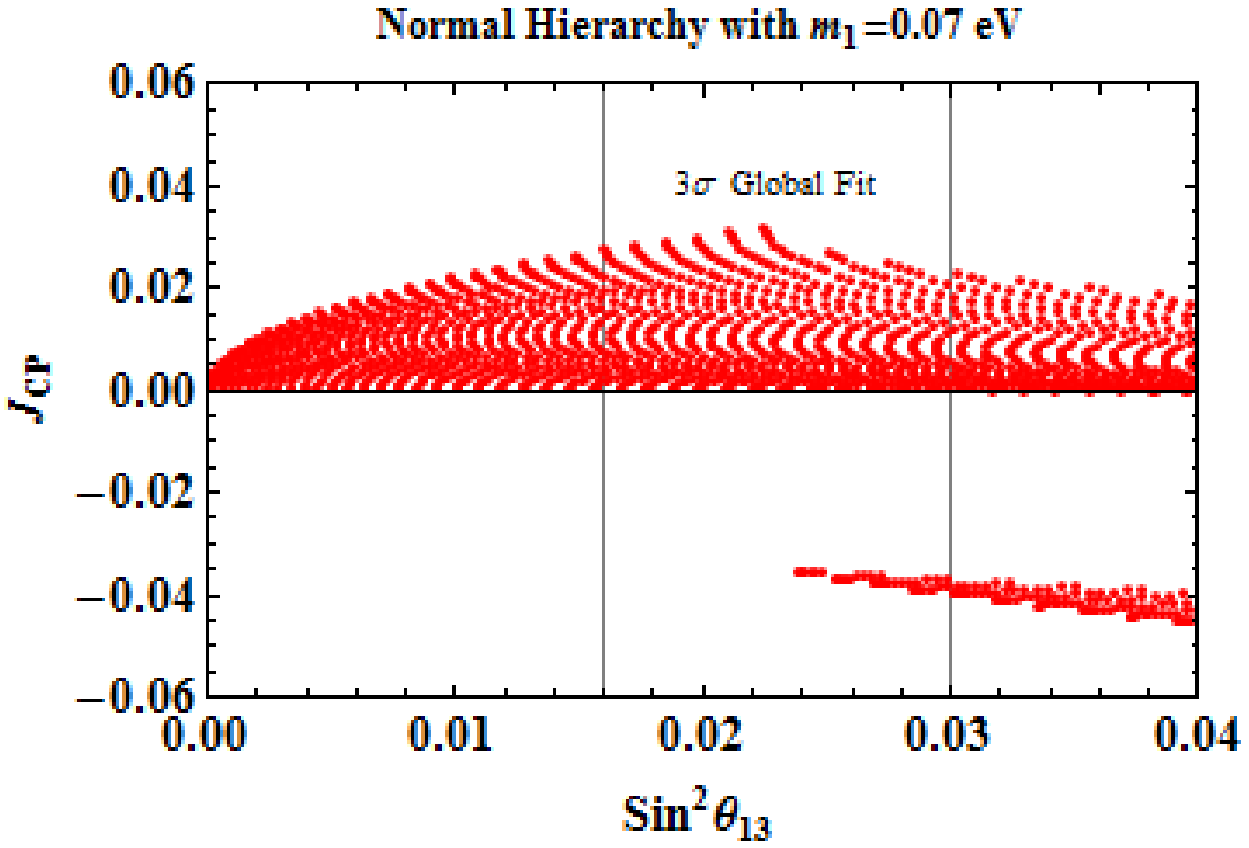} \\
\includegraphics[width=0.5\textwidth]{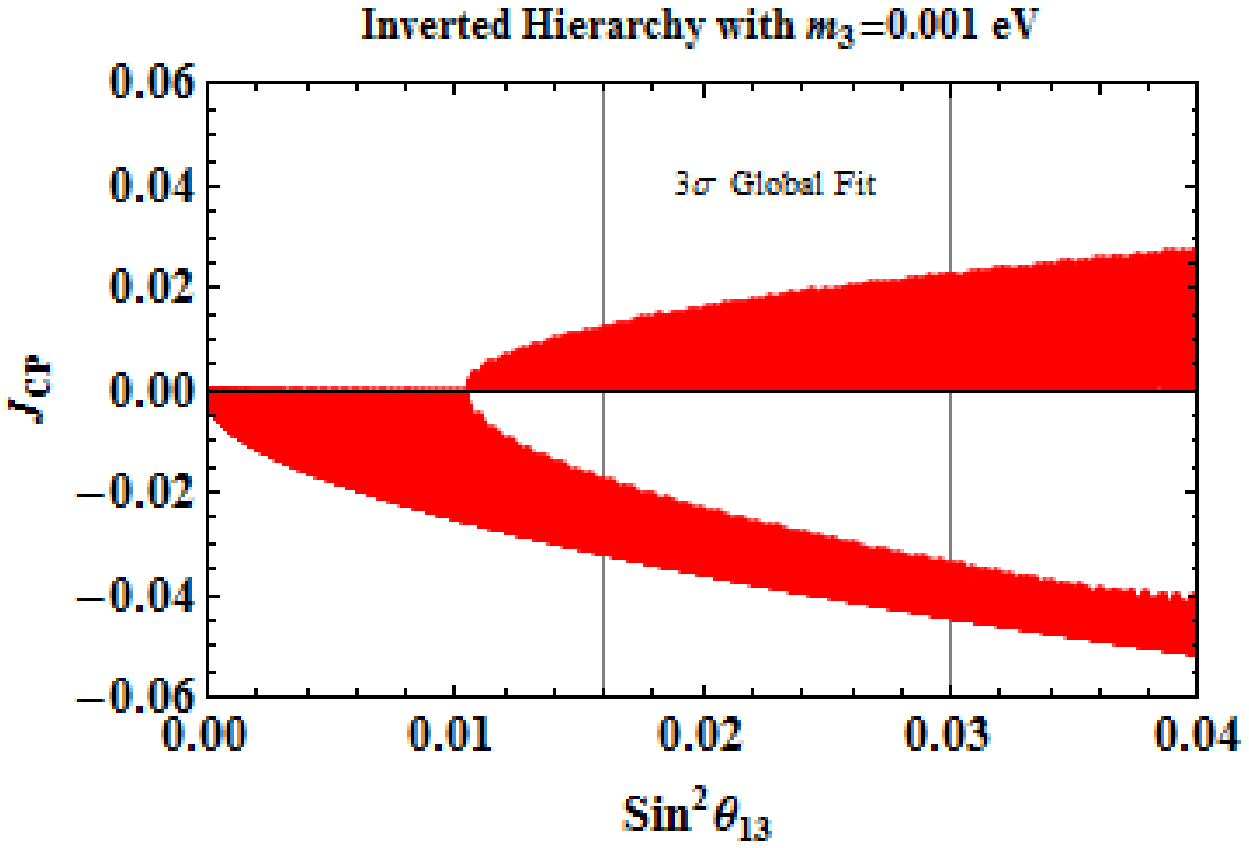} &
\includegraphics[width=0.5\textwidth]{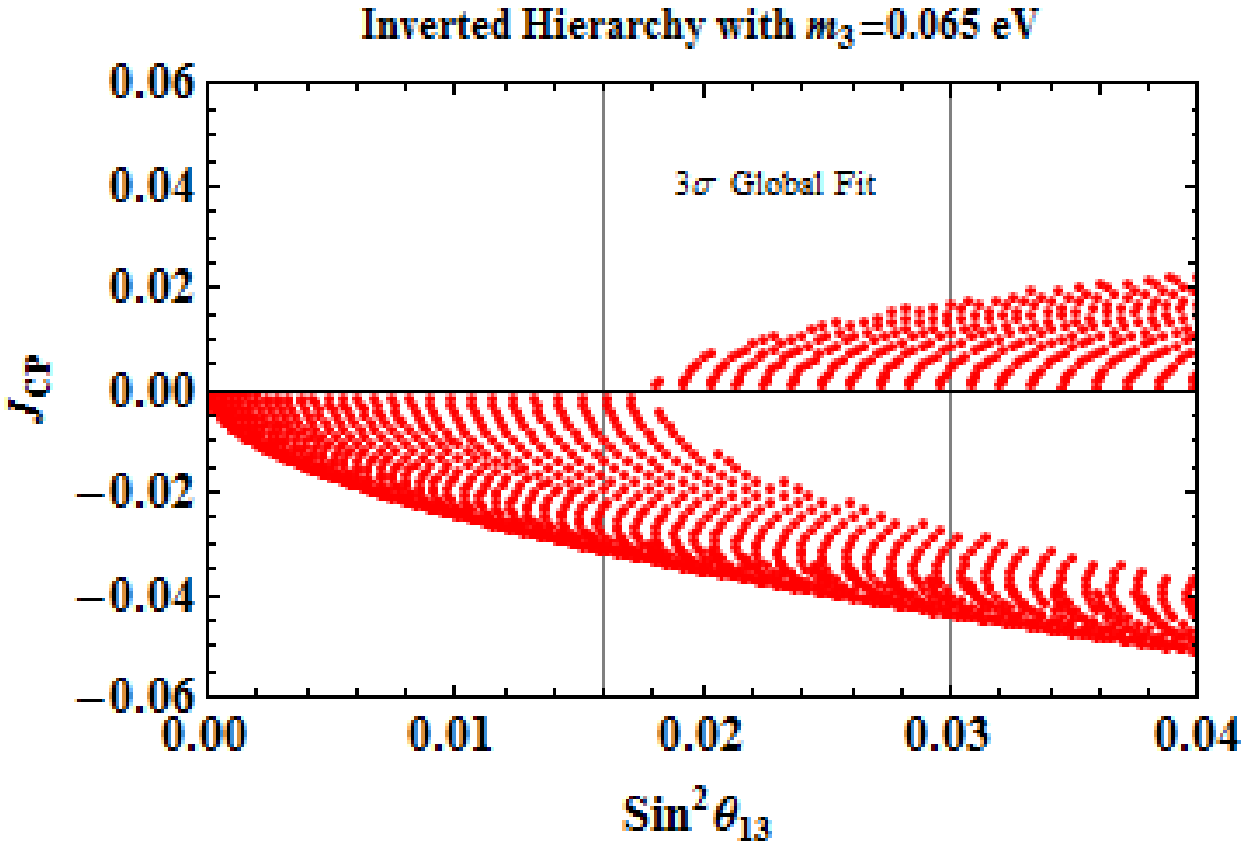}
\end{array}$
\end{center}
\caption{Variation of $J_{CP}$ with $\sin^2{\theta_{13}}$}
\label{fig7}
\end{figure}
\begin{figure}[h]
\begin{center}
$
\begin{array}{cc}
\includegraphics[width=0.5\textwidth]{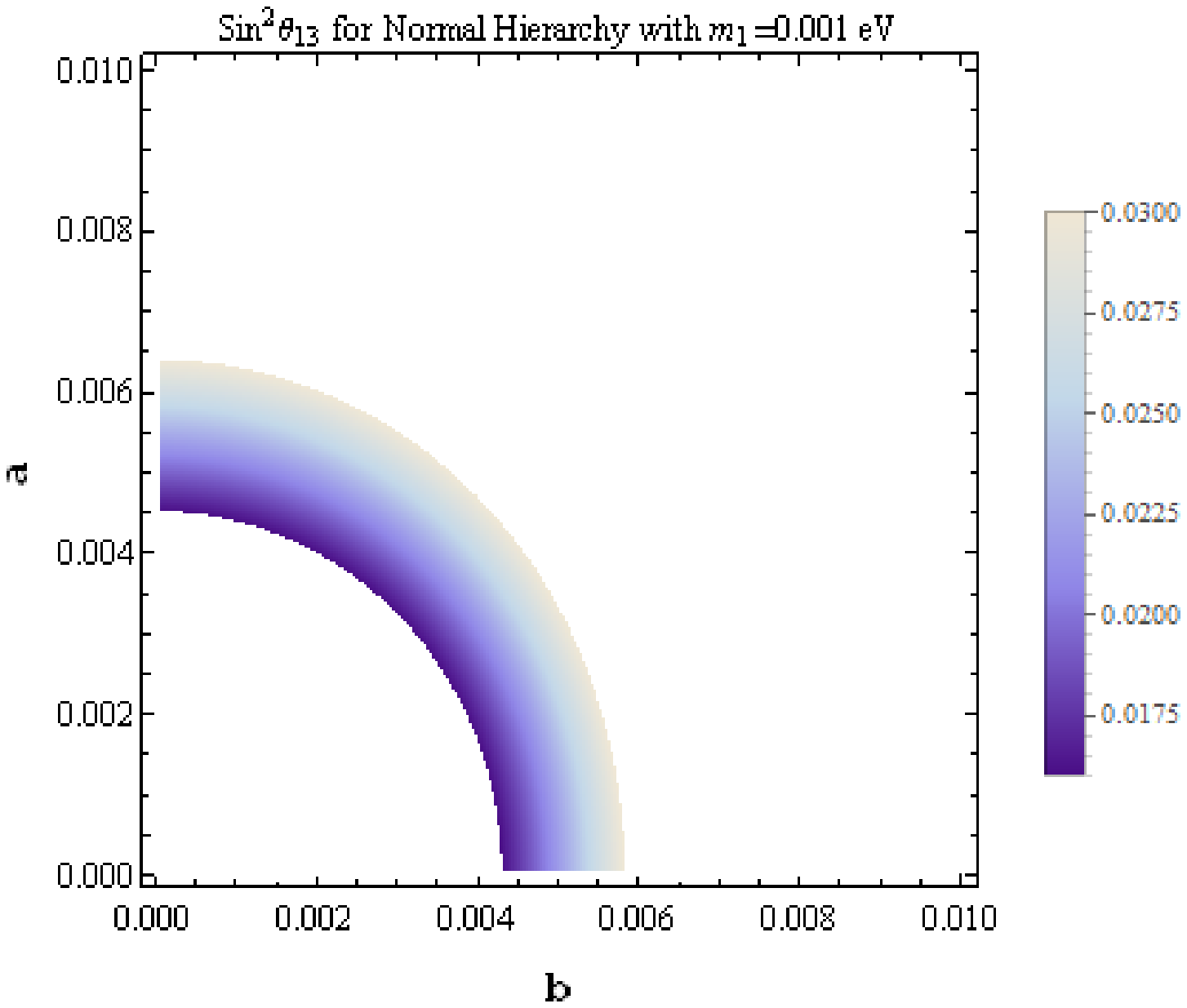} &
\includegraphics[width=0.5\textwidth]{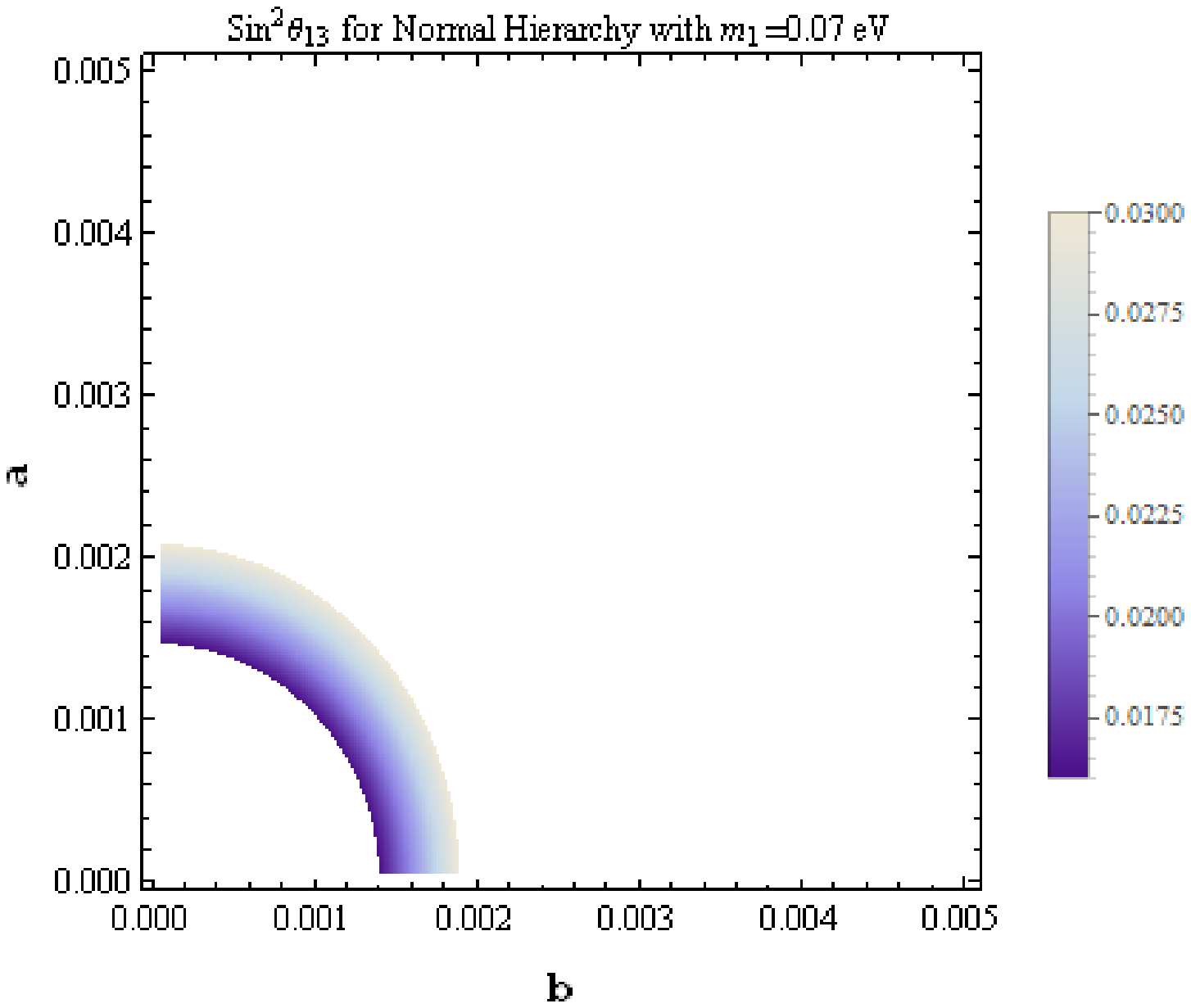} \\
\includegraphics[width=0.5\textwidth]{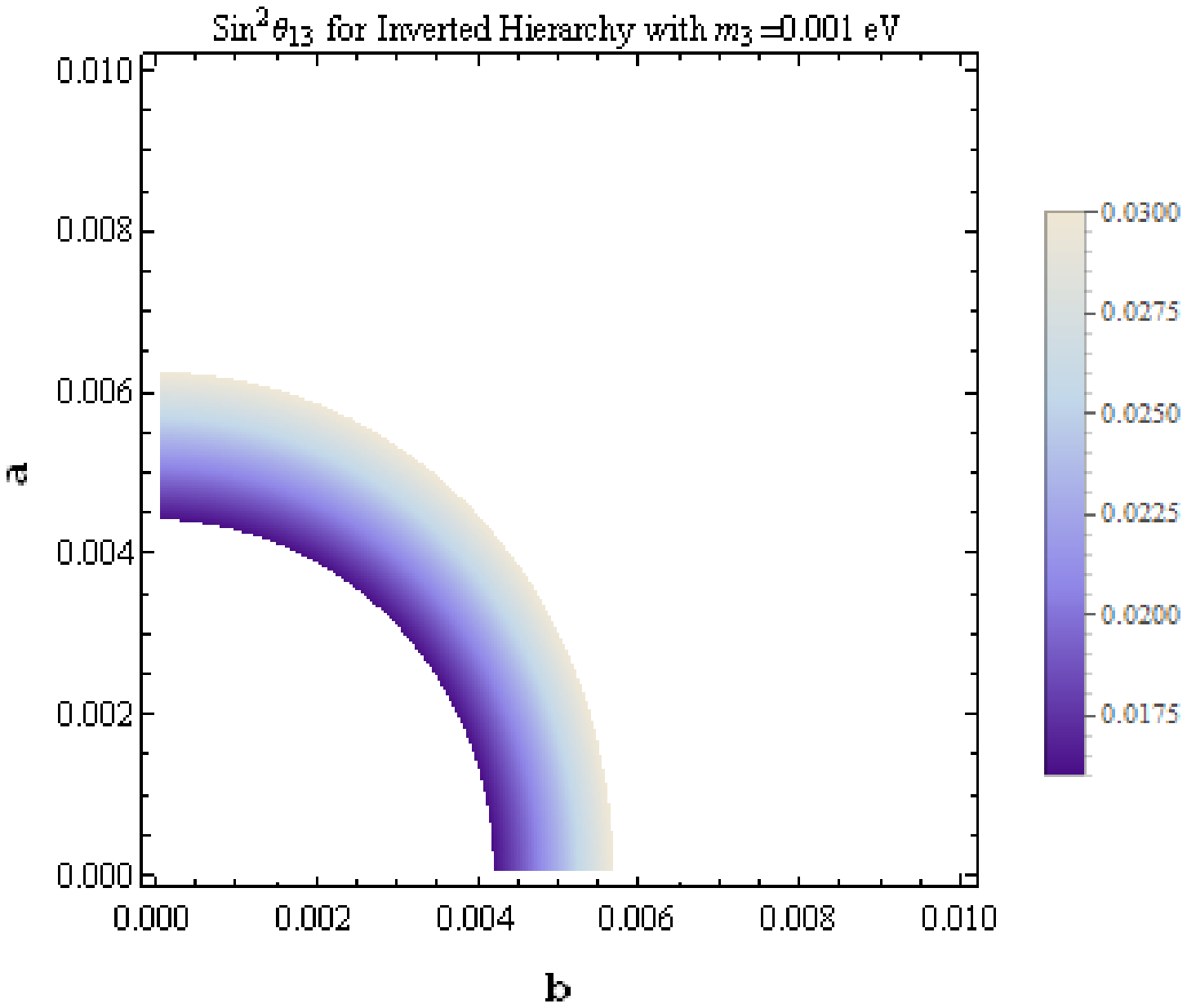} &
\includegraphics[width=0.5\textwidth]{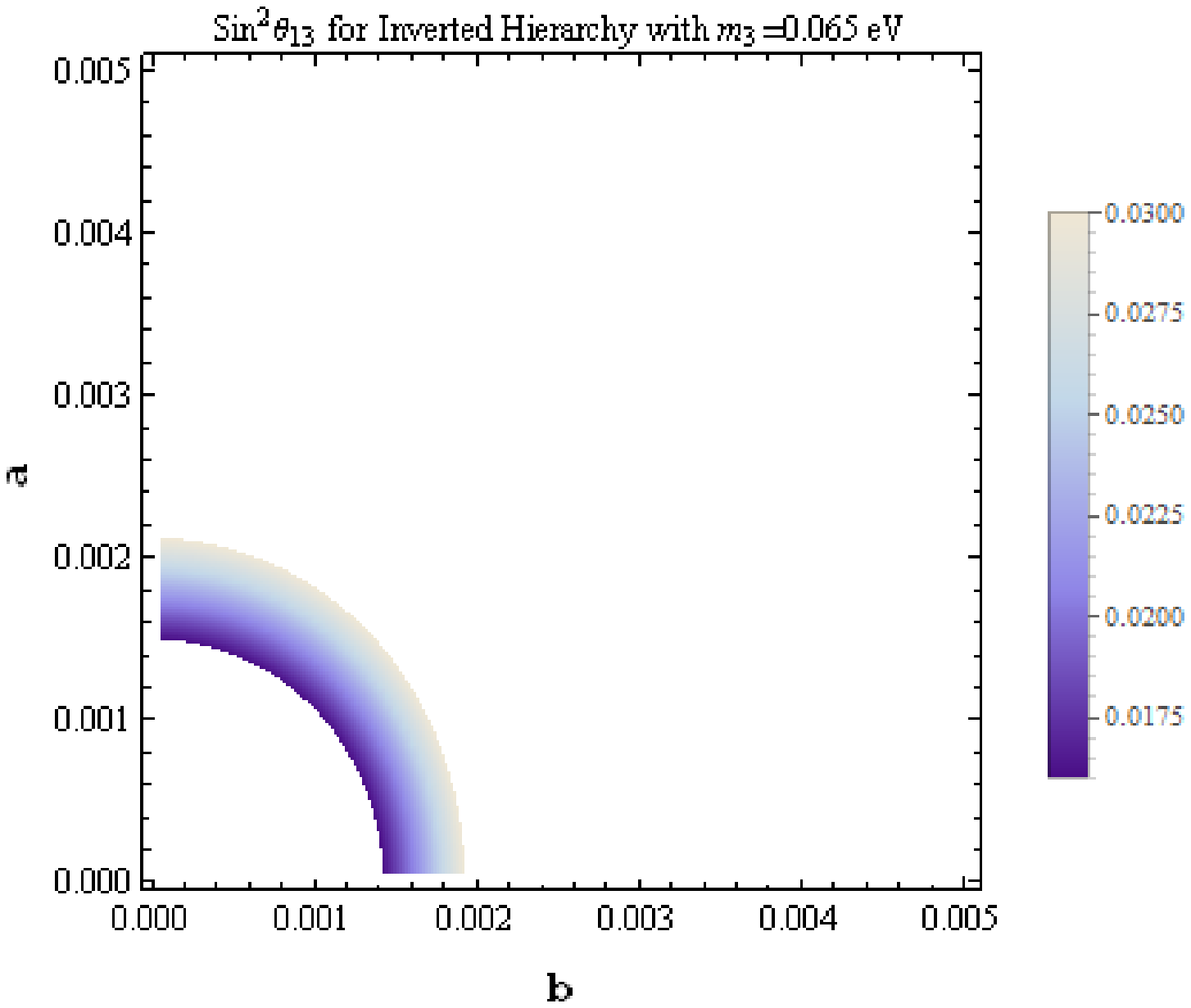}
\end{array}$
\end{center}
\caption{Density plot showing the strength of type II perturbation required to produce $\sin^2{\theta_{13}}$ in the $3\sigma$ range}
\label{fig8}
\end{figure}
\begin{figure}
\begin{center}
\includegraphics[width=0.5\textwidth]{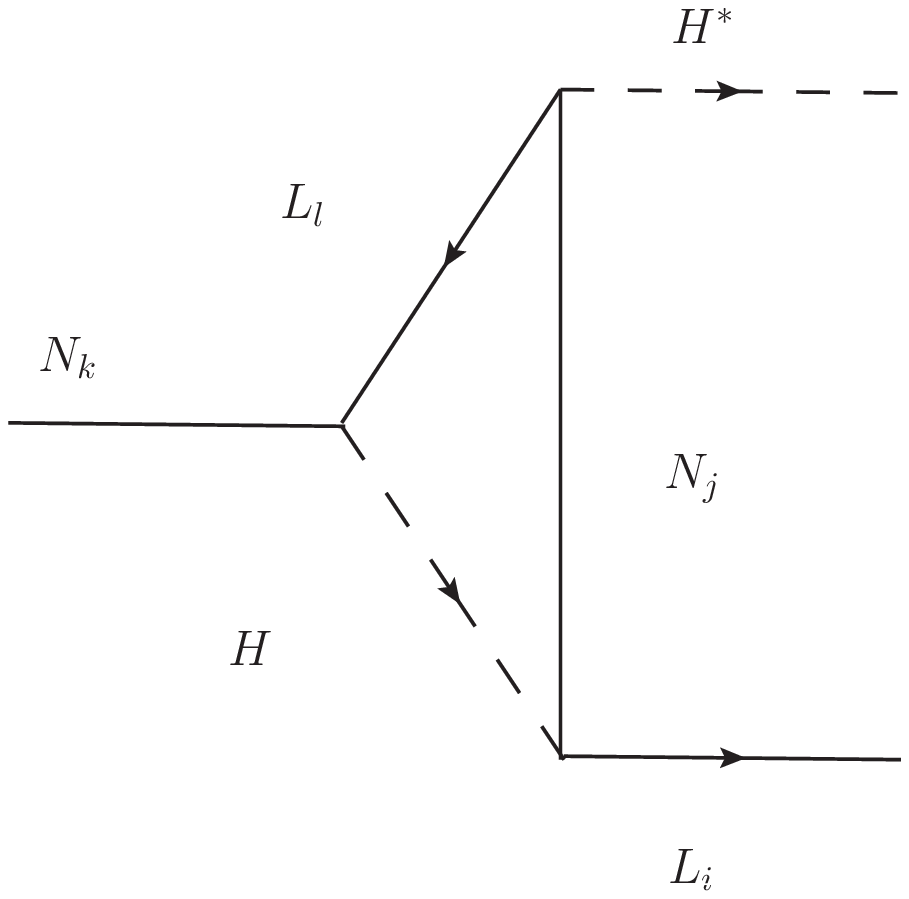}\end{center}
\caption{Right handed neutrino decay}
\label{fig9}
\end{figure}
\begin{figure}
\begin{center}
\includegraphics[width=0.5\textwidth]{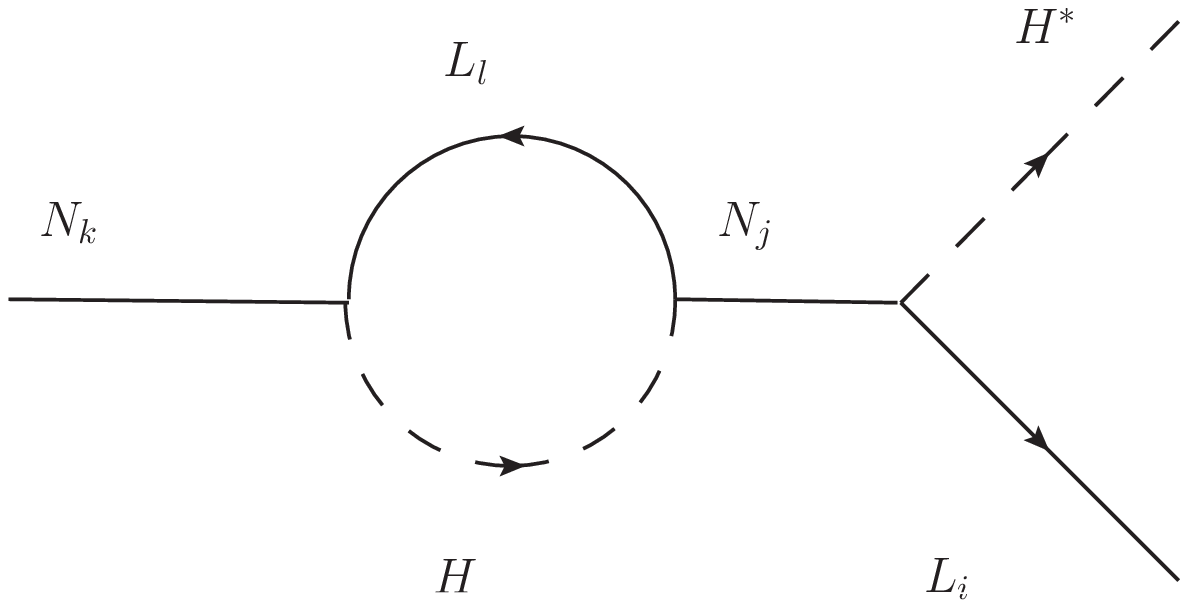}
\end{center}
\caption{Right handed neutrino decay}
\label{fig10}
\end{figure}
\begin{figure}
\begin{center}
\includegraphics[width=0.5\textwidth]{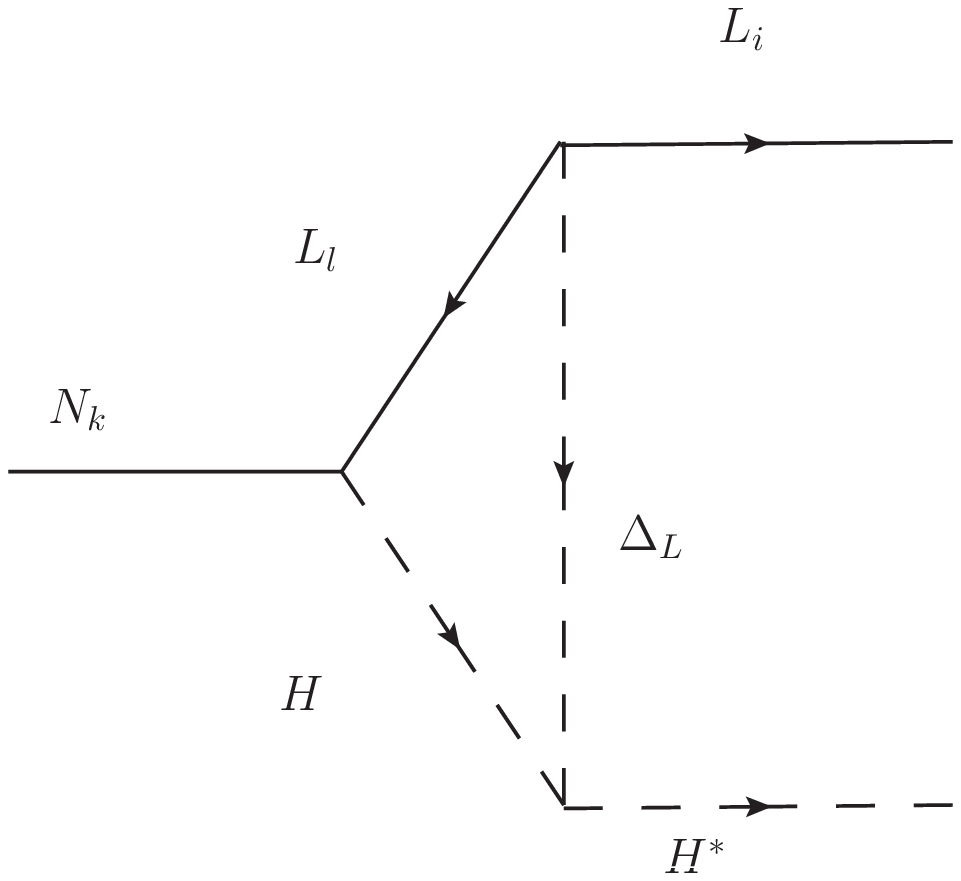}
\end{center}
\caption{Right handed neutrino decay}
\label{fig11}
\end{figure}

\begin{figure}[h]
\begin{center}
$
\begin{array}{cc}
\includegraphics[width=0.5\textwidth]{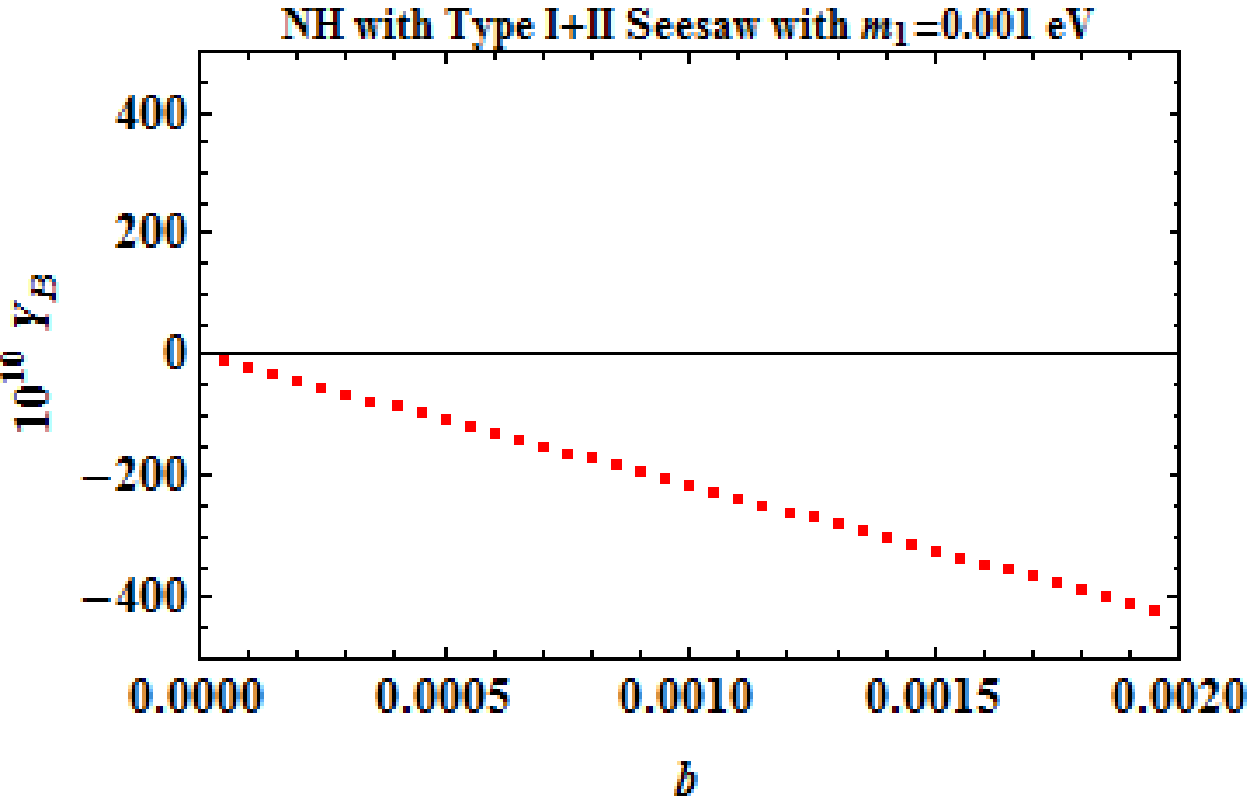} &
\includegraphics[width=0.5\textwidth]{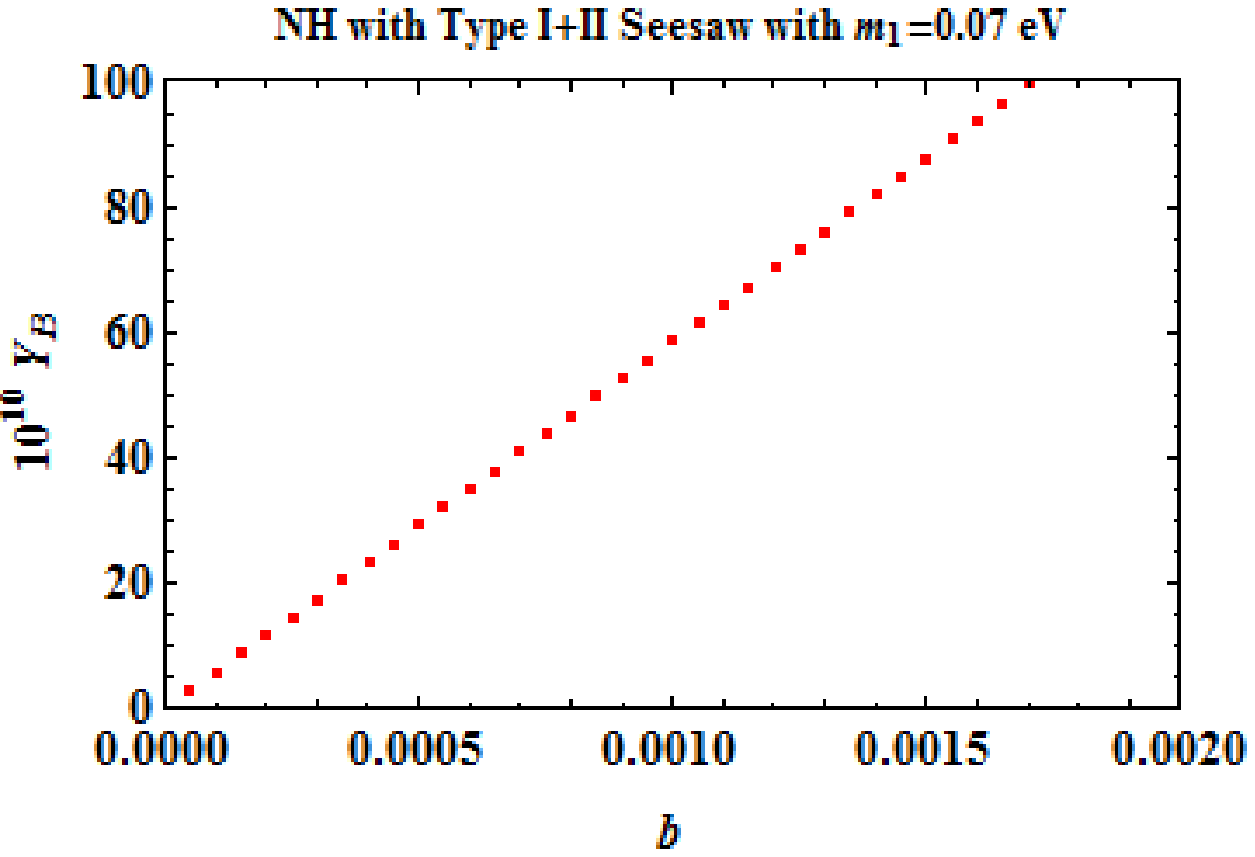} \\
\includegraphics[width=0.5\textwidth]{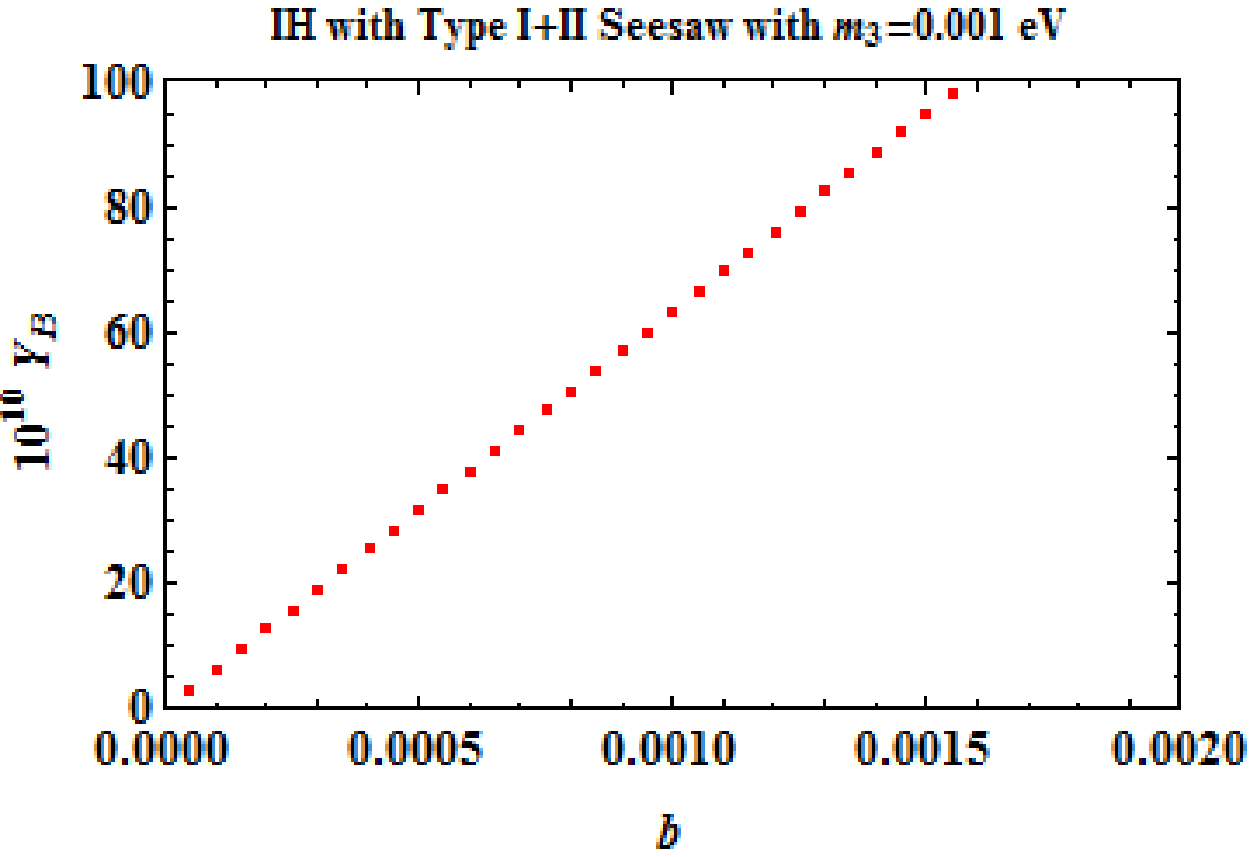} &
\includegraphics[width=0.5\textwidth]{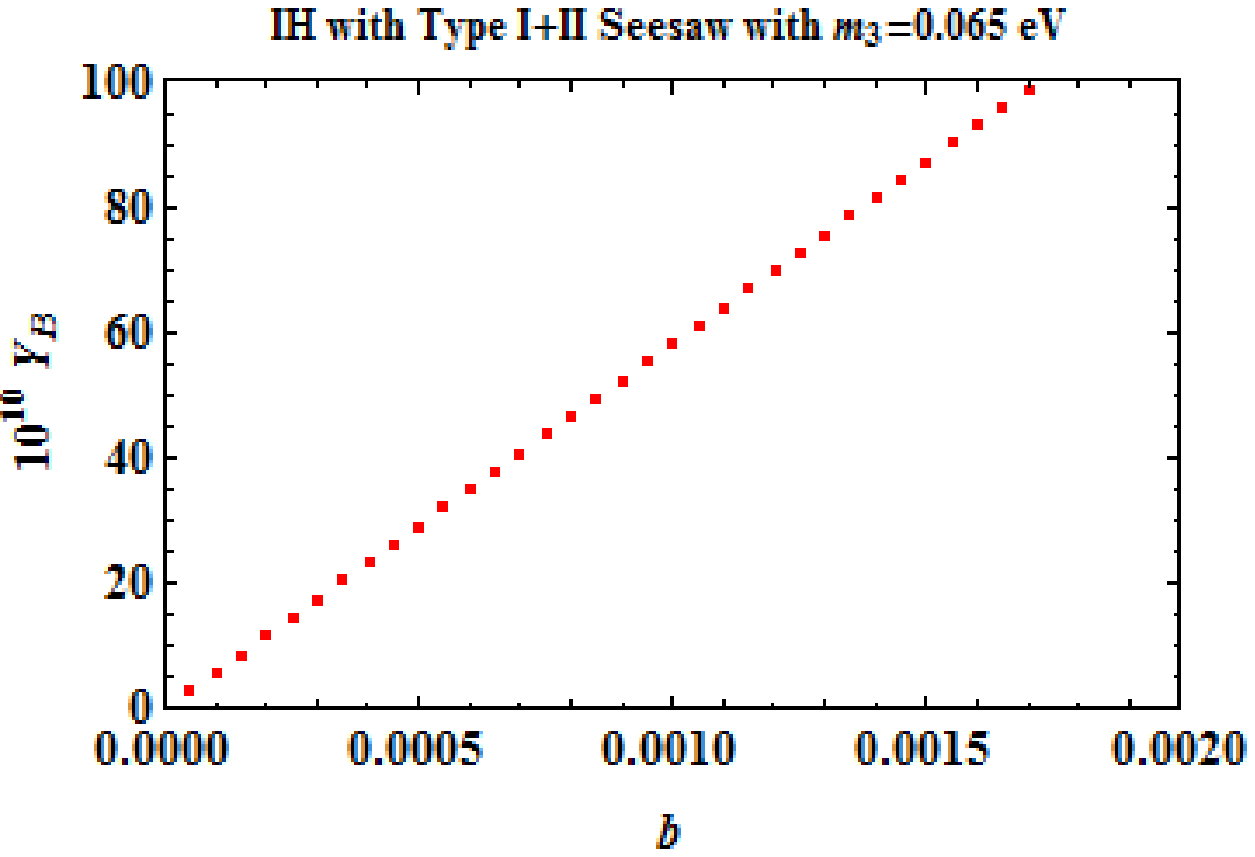}
\end{array}$
\end{center}
\caption{Variation of baryon to photon ratio with $b = \text{Im}(w)$, the imaginary part of the type II seesaw term}
\label{fig12}
\end{figure}
\begin{figure}[h]
\begin{center}
$
\begin{array}{cc}
\includegraphics[width=0.5\textwidth]{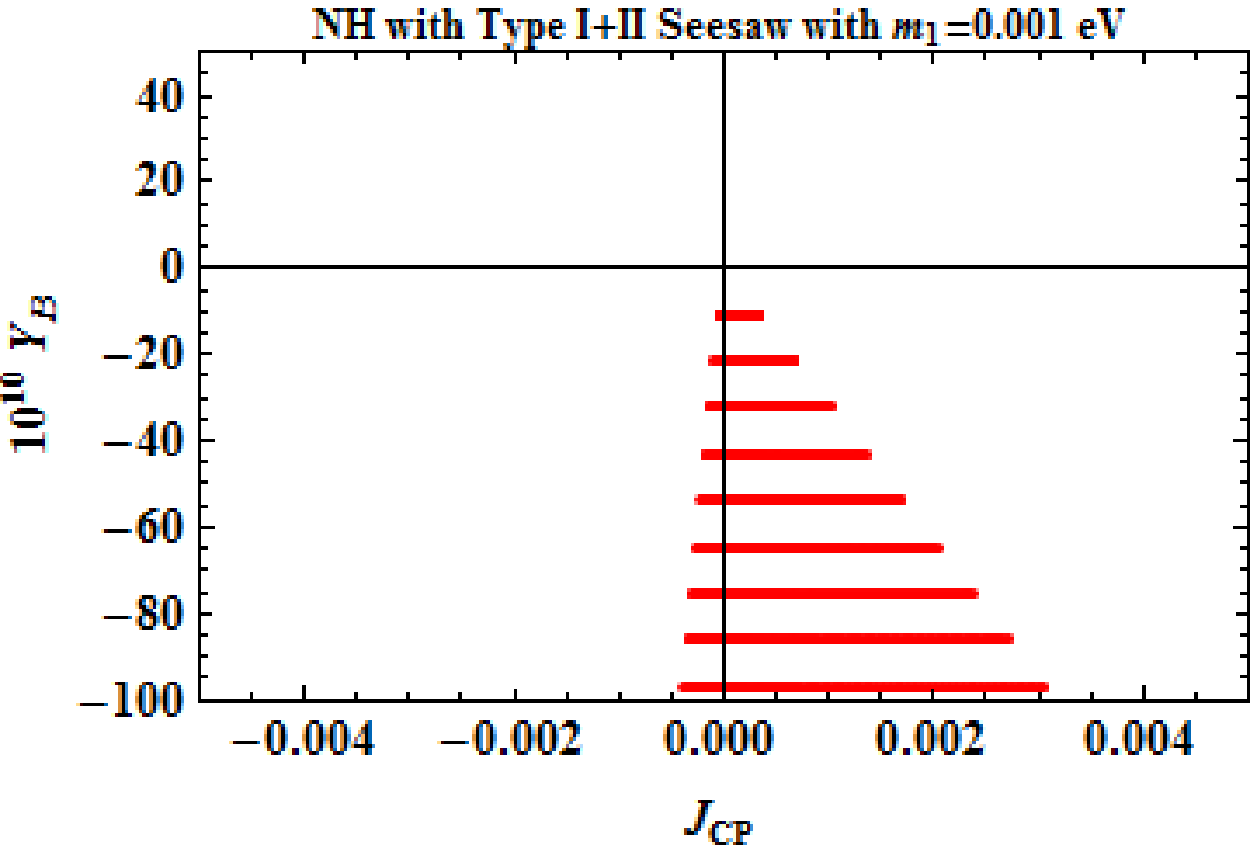} &
\includegraphics[width=0.5\textwidth]{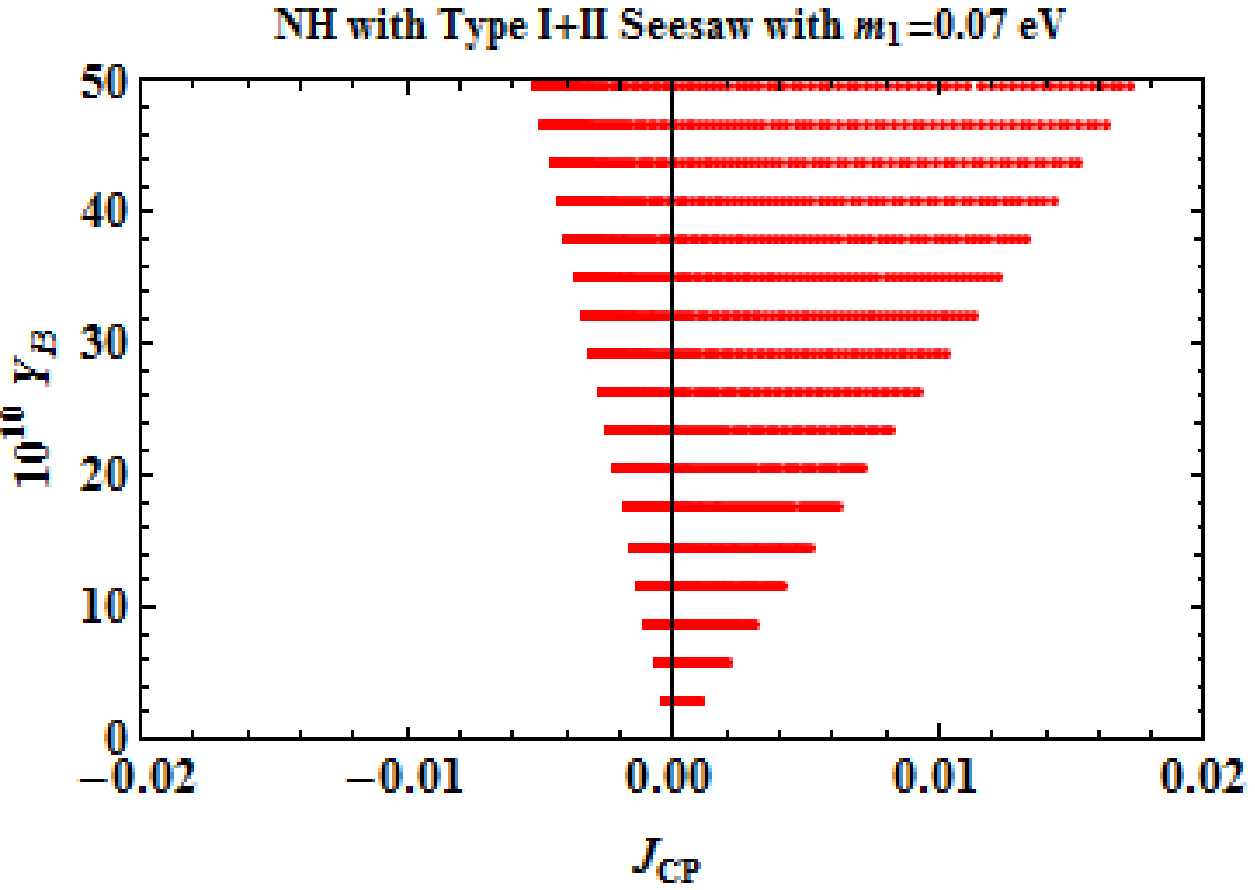} \\
\includegraphics[width=0.5\textwidth]{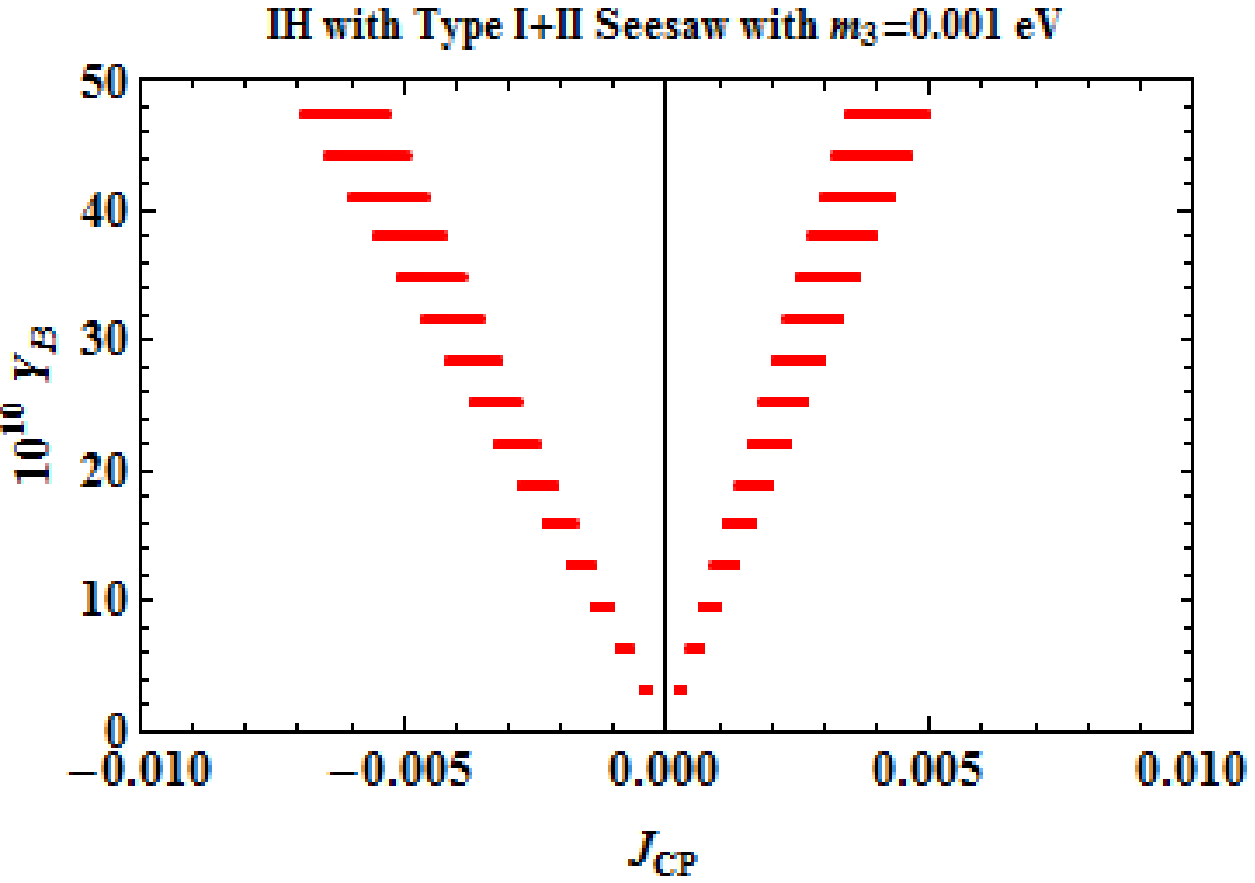} &
\includegraphics[width=0.5\textwidth]{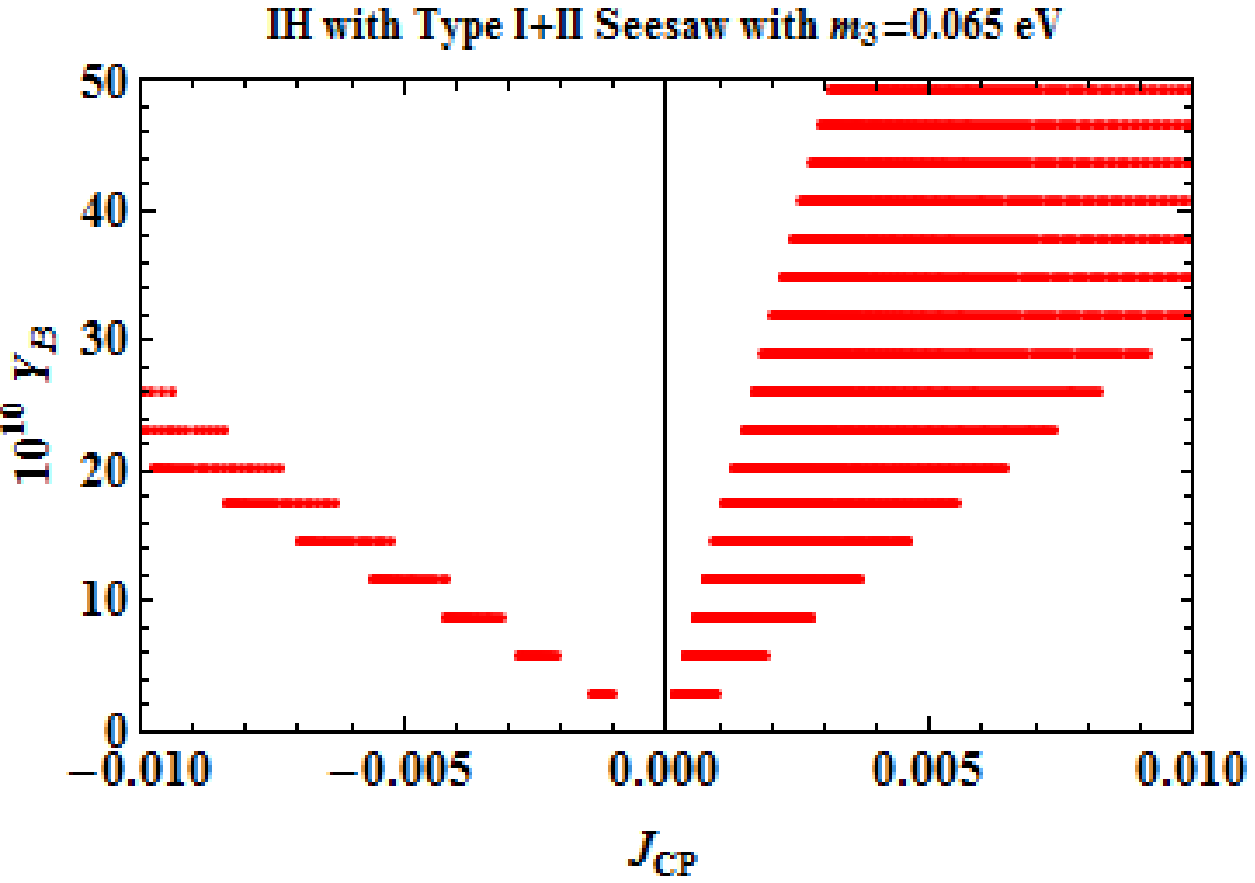}
\end{array}$
\end{center}
\caption{Variation of baryon to photon ratio with $J_{CP}$}
\label{fig13}
\end{figure}

\section{Leptogenesis}
\label{sec:lepto}
As stressed earlier, leptogenesis is a novel mechanism to account for the baryon asymmetry of the Universe by creating an asymmetry in the leptonic sector first, which subsequently gets converted into baryon asymmetry through $B+L$ violating sphaleron processes during electroweak phase transition. Since quark sector CP violation is not sufficient for producing observed baryon asymmetry, a framework explaining non-zero $\theta_{13}$ and leptonic CP phase could not only give a better picture of leptonic flavor structure, but also the origin of matter-antimatter asymmetry. 

In a model with both type I and type II seesaw mechanisms at work, there are two possible sources of lepton asymmetry: either the CP violating decay of right handed neutrino or that of scalar triplet. In our work we are considering dominant type I and sub-dominant type II seesaw which naturally point towards heavier triplet than right handed neutrinos. For simplicity we consider only the right handed neutrino decay as a source of lepton asymmetry and neglect the contribution coming from triplet decay. The lepton asymmetry from the decay of right handed neutrino into leptons and Higgs scalar is given by
\begin{equation}
\epsilon_{N_k} = \sum_i \frac{\Gamma(N_k \rightarrow L_i +H^*)-\Gamma (N_k \rightarrow \bar{L_i}+H)}{\Gamma(N_k \rightarrow L_i +H^*)+\Gamma (N_k \rightarrow \bar{L_i}+H)}
\end{equation}
In a hierarchical pattern for right handed neutrinos $M_{2,3} \gg M_1$, it is sufficient to consider the lepton asymmetry produced by the decay of lightest right handed neutrino $N_1$ decay. In a type I seesaw framework where the particle content is just the standard model with three additional right handed neutrinos, the lepton asymmetry generated through the decay processes shown in figure \ref{fig9} and \ref{fig10} can be estimated as (for a review please refer to \cite{davidsonPR}):
\begin{eqnarray}
\epsilon^{\alpha}_1 &=& \frac{1}{8\pi v^2}\frac{1}{(m^{\dagger}_{LR}m_{LR})_{11}} \sum_{j=2,3} \text{Im}[(m^*_{LR})_{\alpha 1}(m^{\dagger}_{LR}m_{LR})_{1j}(m_{LR})_{\alpha j}]g(x_j) \nonumber \\
&& + \frac{1}{8\pi v^2}\frac{1}{(m^{\dagger}_{LR}m_{LR})_{11}} \sum_{j=2,3} \text{Im}[(m^*_{LR})_{\alpha 1}(m^{\dagger}_{LR}m_{LR})_{j1}(m_{LR})_{\alpha j}]\frac{1}{1-x_j}
\label{eps1}
\end{eqnarray}
where
$$ g(x) = \sqrt{x} \left ( 1+\frac{1}{1-x}-(1+x)\text{ln}\frac{1+x}{x} \right) $$
and $x_j = M^2_j/M^2_1$. The second term in the expression for $\epsilon^{\alpha}_1$ above vanishes when summed over all the flavors $\alpha = e, \mu, \tau$. The sum over flavors is given by
\begin{equation}
\epsilon_1 = \frac{1}{8\pi v^2}\frac{1}{(m^{\dagger}_{LR}m_{LR})_{11}}\sum_{j=2,3} \text{Im}[(m^{\dagger}_{LR}m_{LR})^2_{1j}]g(x_j)
\label{eps2}
\end{equation}
It is important to note that a non-vanishing lepton asymmetry is generated only when the right handed neutrino decay is out of equilibrium. Otherwise both the forward and the backwards processes will happen at the same rate resulting in a vanishing asymmetry. Departure from equilibrium can be estimated by comparing the interaction rate with the expansion rate of the Universe. At very high temperatures $(T \geq 10^{12} \text{GeV})$ all charged lepton flavors are out of equilibrium and hence all of them behave similarly. However at temperatures $ T < 10^{12}$ GeV $(T < 10^9 \text{GeV})$, interactions involving tau (muon) Yukawa couplings enter equilibrium and flavor effects become important \cite{leptoflavor}. Taking these flavor effects into account, the final baryon asymmetry is given by 
\begin{equation}
Y^{2 flavor}_B = \frac{-12}{37g^*}[\epsilon_2 \eta\left (\frac{417}{589}\tilde{m_2} \right)+\epsilon^{\tau}_1\eta\left (\frac{390}{589}\tilde{m_{\tau}}\right )] \nonumber
\end{equation}
\begin{equation}
Y^{3 flavor}_B = \frac{-12}{37g^*}[\epsilon_e \eta\left (\frac{151}{179}\tilde{m_e}\right)+ \epsilon^{\mu}_1 \eta\left (\frac{344}{537}\tilde{m_{\mu}}\right)+\epsilon^{\tau}_1\eta\left (\frac{344}{537}\tilde{m_{\tau}} \right )] \nonumber
\end{equation}
where $\epsilon_2 = \epsilon^e_1 + \epsilon^{\mu}_1, \tilde{m_2} = \tilde{m_e}+\tilde{m_{\mu}}, \tilde{m_{\alpha}} = \frac{(m^*_{LR})_{\alpha 1} (m_{LR})_{\alpha 1}}{M_1}$ and the factor $g_*$ is the effective number of relativistic degrees of freedom at $T=M_1$ and is approximately $110$. The function $\eta$ is given by 
$$ \eta (\tilde{m_{\alpha}}) = \left [\left ( \frac{\tilde{m_{\alpha}}}{8.25 \times 10^{-3} \text{eV}} \right )^{-1}+ \left ( \frac{0.2\times 10^{-3} \text{eV}}{\tilde{m_{\alpha}}} \right )^{-1.16} \right ]^{-1} $$
In the presence of an additional scalar triplet, the right handed neutrino can also decay through a virtual triplet as shown in figure \ref{fig11}. The contribution of this diagram to lepton asymmetry can be estimated as \cite{tripletlepto}
\begin{equation} 
\epsilon^{\alpha}_{\Delta 1}=-\frac{M_1}{8\pi v^2} \frac{\sum_{j=2,3} \text{Im} [(m_{LR})_{1j}(m_{LR})_{1\alpha}(M^{II*}_{\nu})_{j\alpha}]}{\sum_{j=2,3} \lvert (m_{LR})_{1j}\rvert^2}
\label{eps3}
\end{equation}
where $M_1 \ll M_{\Delta}$ is assumed which is natural in a model with dominant type I and sub-dominant type II seesaw mechanisms. 

For the calculation of baryon asymmetry, we go to the basis where the right handed Majorana neutrino mass matrix is diagonal
\begin{equation}
U^*_R M_{RR} U^{\dagger}_R = \text{diag}(M_1, M_2, M_3)
\label{mrrdiag}
\end{equation}
In this diagonal $M_{RR}$ basis, the Dirac neutrino mass matrix also changes to 
\begin{equation}
m_{LR} = m^d_{LR} U_R
\label{mlrdiag}
\end{equation}
where $m^d_{LR}$ is the assumed diagonal choice of the Dirac neutrino mass matrix in our calculation. This is the $m_{LR}$ (in the basis where right handed neutrino mass matrix is diagonal) that appears in the expression for lepton asymmetry in equations (\ref{eps1}), (\ref{eps2}), (\ref{eps3}). If we choose the Dirac neutrino mass matrix to be real and diagonal, say charged lepton (CL) type or up quark (UQ) type, the right handed neutrino mass matrix $M_{RR}$ constructed from 
$$M_{RR} = m^T_{LR}m^{-1}_{LL}m_{LR}$$
also remains real owing to the fact that the neutrino mass matrix $m_{LL}$ coming from type I seesaw is of $\mu-\tau$ symmetric type giving rise to TBM mixing. Since $M_{RR}$ is real, its diagonalizing matrix is real and hence the Dirac neutrino mass matrix in the diagonal $M_{RR}$ basis (\ref{mlrdiag}) is also real. This gives rise to vanishing lepton asymmetry originating from right handed neutrino decay (\ref{eps1}), (\ref{eps2}) within a type I seesaw framework. However, in the presence of type II contribution, the lepton asymmetry originating from right handed neutrino decay through a virtual Higgs triplet (\ref{eps3}) can be non-zero if the type II seesaw term $M^{II}_{\nu}$ contains non-extremal CP phases. This is what we pursue in the remaining part of our paper: type II seesaw as common origin of non-zero $\theta_{13}$ as well as lepton asymmetry. It should be noted that the CP violating decay of scalar triplet can also give additional contribution to lepton asymmetry. However, in our analysis we neglect these extra contributions coming from triplet decay. Some recent studies on flavor effects in scalar triplet leptogenesis have been done in \cite{tripletflav}.

\section{Numerical Analysis and Results}
\label{numeric}
For our numerical analysis, we adopt the minimal structure (\ref{matrix3}) of the type II seesaw term. We first numerically fit the leading order $\mu-\tau$ symmetric neutrino mass matrix (\ref{matrix1}) by taking the central values of the global fit neutrino oscillation data \cite{schwetz12}. We also incorporate the cosmological upper bound on the sum of absolute neutrino masses \cite{Planck13} reported by the Planck collaboration recently. For normal hierarchy, the diagonal mass matrix of the light neutrinos can be written 
 as $m_{\text{diag}} = \text{diag}(m_1, \sqrt{m^2_1+\Delta m_{21}^2}, \sqrt{m_1^2+\Delta m_{31}^2})$ whereas for inverted hierarchy 
 it can be written as $m_{\text{diag}} = \text{diag}(\sqrt{m_3^2+\Delta m_{23}^2-\Delta m_{21}^2}, \sqrt{m_3^2+\Delta m_{23}^2}, m_3)$. 
 We choose two possible values of the lightest mass eigenstate $m_1, m_3$ for normal and inverted hierarchies respectively. First 
 we choose $m_{\text{lightest}}$ as large as possible such that the sum of the absolute neutrino masses fall just below the cosmological 
 upper bound. For normal and inverted hierarchies, this turns out to be $0.07$ eV and $0.065$ eV respectively. Then we allow moderate 
 hierarchy to exist between the mass eigenvalues and choose the lightest mass eigenvalue to be $0.001$ eV to study the possible changes 
 in our analysis and results. The parametrization for all these possible cases are shown in table\, \ref{table:results1}.

After fitting the type I seesaw contribution to neutrino mass with experimental data, we introduce the type II seesaw contribution as a perturbation to the TBM neutrino mixing. We allow the perturbation to be complex ($w = a+ib$) such that it can account for non-zero $\theta_{13}$ and Dirac CP phase $\delta_{CP}$ simultaneously. We compute the predictions for neutrino parameters by varying $a, b$ and show the results as a function of $\sin^2{\theta_{13}}$ in figure \ref{fig1}, \ref{fig2}, \ref{fig3} and \ref{fig4}. All the neutrino parameters lie well within $3\sigma$ range except for $\sin^2{\theta_{12}}$ which lie very close to the upper limit of $3\sigma$ range as can be seen from figure \ref{fig3}. As seen from figure \ref{fig4}, the angle $\theta_{23}$ lies in first and second octant for inverted and normal hierarchies respectively. We also compute the sum of absolute neutrino masses $\sum_i \lvert m_i \rvert$ which lies below the Planck bound as can be seen from figure \ref{fig5}. We also compute the effective neutrino mass $m_{ee} = \lvert \sum_i U^2_{ei} m_i \rvert$ which can play a role in neutrino-less double beta decay experiments and show its variation with $\theta_{13}$ in figure \ref{fig6}. To see the variation of Dirac CP phase, we compute Jarlskog's rephasing invariant CP violation measure $J_{CP}$ \cite{jarlskog} given by
$$J_{CP} = \text{Im}[U^*_{e1}U^*_{\mu 3}U_{e3}U_{\mu 1}]$$
and plot it as a function of $\sin^2{\theta_{13}}$ in figure \ref{fig7}. We finally show the density plot of $\sin^2{\theta_{13}}$ as a function of the perturbations $a,b$ to show the required strength of perturbations in order to generate non-zero $\theta_{13}$ in the correct $3\sigma$ range. It is seen from figure \ref{fig8} that for higher values of $m_{\text{lightest}}$, we require a lower strength of the type II seesaw term to 
give rise to the desired $\theta_{13}$. For $m_{\text{lightest}}=0.065, 0.07$ eV, one can see from figure \ref{fig5} that $a,b \sim 
0.0015-002 \; \text{eV} \Rightarrow \frac{\mu_{\Delta H}\langle \phi^0 \rangle^2}{M^2_{\Delta}} = 0.0015-0.002$ eV. Taking  $\langle \phi^0 \rangle = 10^2$ GeV and $\mu_{\Delta H} \sim M_{\Delta}$ we get a bound on $M_{\Delta} \sim 10^{16}$ GeV to get the desired type II seesaw strength. Similarly, for $m_{\text{lightest}} = 0.001$ eV, one gets a bound $a,b \sim 0.00425-0.00612$ eV which also puts a similar constraint on $M_{\Delta}$. This constraint $M_{\Delta} \sim 10^{16}$ GeV could point towards the origin of type II seesaw or the Higgs triplet from a grand unified theory where the scale of unification is typically of the order $10^{16}$ GeV.

We then compute the predictions for baryon to photon ratio following the procedure discussed in the previous section. We consider a scenario where only the type II seesaw contribution gives rise to lepton asymmetry given by equation (\ref{eps3}) whereas type I seesaw contribution does not give rise to any non-trivial CP violating phase in the mass matrix resulting in a vanishing contribution to lepton asymmetry. To make use of equation (\ref{eps3}), we have to assume a specific form of the Dirac neutrino mass matrix. It is quite generic to choose the structure as
\begin{equation}
m_{LR}=\left(\begin{array}{ccc}
\lambda^m & 0 & 0\\
0 & \lambda^n & 0 \\
0 & 0 & 1
\end{array}\right)m_f
\label{mLR1}
\end{equation}
where $m_f$ corresponds to $m_\tau \tan{\beta}$ for $(m, n) = (6, 2), \; \tan{\beta} = 40$ in case of charged lepton (CL) and $m_t$ for $(m, n) = (8, 4)$ in the case of up-quarks (UQ)\cite{dm,mkd}. $\lambda = 0.22$ is the standard Wolfenstein parameter. Using this input we calculate the right-handed Majorana neutrino matrix $M_{RR}$ using the inverse type I seesaw formula 
\begin{equation}
M_{RR}=m_{LR}^Tm_{LL}^{-1}m_{LR}
\end{equation}
where $m_{LL}$ is given by equation (\ref{matrix1}) with $x,y,z$ values given in table \ref{table:results1}. It turns out that for either CL or UQ type $m_{LR}$, the lightest right handed neutrino turns out to be heavier than $10^{12}$ GeV. In this regime, lepton flavor effects do not play any role in leptogenesis. As mentioned before, for lightest right handed neutrino mass within $10^9 \; \text{GeV} < M_1 < 10^{12} \; \text{GeV}$, tau lepton Yukawa interactions enter thermal equilibrium and hence flavor effects become very important. We consider this interesting scenario where two flavor leptogenesis is applicable. To keep the lightest right handed neutrino mass in this two flavor regime, we need to choose the Dirac neutrino mass matrix with appropriate diagonal entries. We take the same value of lightest active neutrino mass as considered before, and vary $(m,n)$ of $m_{LR}$ given in equation (\ref{mLR1}) to calculate the right handed neutrino masses. We find that the choice $(m,n)=(3,1)$ keeps the lightest right handed neutrino in the two flavor regime and also maintain a sizable hierarchy (two orders of magnitudes) between the right handed neutrinos. Other choices of $(m,n)$ lower the hierarchy and hence not considered in our work. We then compute the lepton asymmetry originating from the decay of lightest right handed neutrino through a virtual Higgs triplet (figure \ref{fig11}) using equation (\ref{eps3}). Since the lepton asymmetry is directly proportional to the type II seesaw term, the final baryon to photon ratio varies linearly with the type II seesaw strength as seen from figure \ref{fig12}. In figure \ref{fig13}, we show the variation of baryon to photon ratio as a function of Jarlskog CP violation parameter $J_{CP}$. This shows the overall dependence of baryon asymmetry on the Dirac CP violating phase. Apart from normal hierarchy with $m_1 = 0.001$ eV, all other cases under consideration give rise to desired baryon asymmetry. It is also seen from figure \ref{fig13}, that to produce the correct baryon asymmetry (\ref{barasym}), the Jarlskog parameter $J_{CP}$ needs to be very small in magnitude. Such a small value of $J_{CP}$ is also consistent with the required value of $\sin^2{\theta_{13}}$ as can be seen from figure \ref{fig7}. In other words, the imaginary part of the type II seesaw term has to be very small $b \sim 0.0001$ to produce the required baryon asymmetry (\ref{barasym}) as seen from figure \ref{fig12}. Such a small value of $b$ can also produce the required $\sin^2{\theta_{13}}$ provided the real part of the type II seesaw term is kept large as depicted by figure \ref{fig8}.

\section{Conclusion}
\label{conclude}
We have studied the possibility of explaining non-zero value of reactor mixing angle as well as Dirac CP phase by introducing the type II seesaw term as a perturbation to the TBM type neutrino matrix derived from type I seesaw mechanism. TBM type neutrino mixing in the context of type I seesaw has been discussed extensively in the literature and for our work we assume this form for type I seesaw skipping the details. The type II seesaw structure is chosen to be minimal which can give rise to the required deviations from TBM mixing by breaking the $\mu-\tau$ symmetry. By varying the strength of this type II seesaw term we compute the predictions for neutrino mixing parameters, sum of absolute neutrino masses, effective neutrino mass, Jarlskog CP parameter and found them to lie in the required range. The value of $\sin^2{\theta_{12}}$ however, lies very close to the $3\sigma$ upper limit and future data may be able to rule out some of these scenarios we discuss in this work. We have done this exercise for both normal and inverted hierarchical neutrino mass spectra as well as two possible values of lightest neutrino mass (one being close to the maximum allowed by cosmological upper bound and one slightly lower). We also constrain the type II seesaw strength and hence the mass of additional Higgs triplet $M_{\Delta}$ such that the desired value of $\theta_{13}$ gets generated through it. We find this constraint on $M_{\Delta}$ to lie very close to the grand unification scale $10^{16}$ GeV. We also find the predictions for baryon to photon ratio by calculating lepton asymmetry from the decay of the lightest right handed neutrino through a virtual Higgs triplet. Other decay channels of right handed neutrino do not contribute to lepton asymmetry as the type II seesaw is the only source of Dirac CP phase in our model. We consider lepton flavor effects into account and show that all the models except the one with normal hierarchy and $m_1 = 0.001$ eV can give rise to the required baryon to photon ratio.

More precise experimental data should be able to shed more light on the viability of these models in giving rise to the correct phenomenology. It should be noted that, our analysis has not considered the effects of Majorana phases (which are currently unconstrained from neutrino oscillation experiments) and are assumed to take only extremal values. A detailed analysis considering wider possibilities of these phases should allow more parameter space which are consistent with the experimental data. We leave such an exercise for future studies.


\end{document}